\documentclass[preprint,review,12pt,a4paper,3p]{elsarticle}
\usepackage[version=3]{mhchem} 
\usepackage[]{geometry}
\usepackage[scriptsize,tight]{subfigure}
\usepackage{graphicx}
\usepackage{multirow}
\usepackage{dcolumn}
\usepackage{bm}
\usepackage{color,soul}
\usepackage[mathlines]{lineno}
\usepackage{hyperref}

\biboptions{longnamesfirst,sort&compress,semicolon}

\begin{document}

\begin{frontmatter}

\title{M\"obius boron--nitride nanobelts interacting with heavy metal
nanoclusters}

\author{C. Aguiar$^{1}$}
\author{N. Dattani$^{2,3}$}
\ead{nike@hpqc.org}
\author{I.~Camps\corref{coric}$^{1,3}$}
\ead{icamps@unifal-mg.edu.br}
\cortext[coric]{Corresponding authors}

\address{$^{1}$Laborat\'orio de Modelagem Computacional - \emph{La}Model,
Instituto de Ci\^{e}ncias Exatas - ICEx. Universidade Federal de Alfenas -
UNIFAL-MG, Alfenas, Minas Gerais, Brasil}
\address{$^{2}$HPQC College, Waterloo, Canada}
\address{$^{3}$HPQC Labs, Waterloo, Canada}

\begin{abstract}
How do nickel, cadmium, and lead nanoclusters interact with boron-nitride and
Mobius-type boron-nitride nanobelts? To answer this question, we used the
semiempirical tight binding framework, as implemented in the xTB software, to
determine the lowest energy geometries, binding energy, complexes stability,
and electronic properties. Our calculations show that heavy metal nanoclusters
favorably bind to both boron-nitride nanobelts, although the interaction is
stronger with the Mobius-type nanobelt. The calculations show that the nickel
nanocluster has the lowest binding energy and the greatest charge transfer with
the nanobelts, followed by the cadmium and lead nanoclusters. During the
simulation time, the molecular dynamic simulation showed that all complexes
were stable at 298 K. Following the nanobelt's symmetry, the frontier orbitals
are distributed homogeneously throughout the structure. This distribution
changed when the nanobelt was twisted to create the Mobius-type nanobelt. The
topological study indicated that the number of bonds between the metal
nanoclusters and the Mobius-type nanobelt doubled and that the bonds formed
with the nickel nanocluster were stronger than those formed with the cadmium
and lead metals. Combining all the results, we conclude that the nickel
nanoclusters are chemisorbed, whereas the cadmium and lead nanoclusters are
physisorbed in both nanobelts.

\end{abstract}


\end{frontmatter}

\section{Introduction}
\label{Sec:Intro}
\newcommand{\sizeA}{3.0cm}

Industrialization and population growth have significantly contributed to the
increase in
pollution~\cite{Mehndiratta-EnvironmentandPollution-2-1-2013,
Zhang-Appl.Mater.Today-7-222-2017}.
 Heavy metal ions, when released into the
environment, can cause numerous diseases such as cancer, hepatitis, spontaneous
abortions, anemia, among
others~\cite{Sardans-SoilSedimentContam.-20-447-2011,
Bali-Int.J.Environ.Sci.Te.-16-249-2018}.
 Lead (Pb), for example, is a
carcinogenic element and its toxicity can cause mental retardation, congenital
defects, brain damage, and death~\cite{Engwa-2019}. Nickel ions (Ni)
can cause skin
diseases~\cite{Moreno-Langmuir-20-8142-2004}. Cadmium (Cd) is not only
carcinogenic but can also cause damage to the
kidneys and respiratory system~\cite{Jaishankar-Interdiscip.Toxicol.-7-60-2014,
Baby-NanoscaleRes.Lett.-14-341-2019}. This is because these metals have
no
biological function, but can accumulate and interfere with metabolism and
physiological processes~\cite{Baby-NanoscaleRes.Lett.-14-341-2019}. Therefore,
prevention, as well as the
identification and treatment of pollutants, are essential for the maintenance
and preservation of the environment. In this sense, nanotechnology can play a
fundamental role in identifying and treating
effluents~\cite{Baby-NanoscaleRes.Lett.-14-341-2019,
Wu-Environ.Pollut.-246-608-2019}.

Nanocarbon materials, due to their large surface area, have mesopores that make
them ideal for removing heavy metal ions through
adsorption~\cite{Gupta-WaterRes.-45-2207-2011,
Burakov-Ecotox.Environ.Safe.-148-702-2018,
Baby-NanoscaleRes.Lett.-14-341-2019}.
Moreover, these nanomaterials can be functionalized with numerous molecules
that can make them specific to adsorb certain
materials~\cite{bastos-Appl.Surf.Sci.-285P-198-2013,
Baby-NanoscaleRes.Lett.-14-341-2019}.
 Carbon nanostructures have
already been reported for use in heavy metal adsorption and are also used in
purifying water contaminated with heavy
metals~\cite{Kaneko-J.Phys.Chem.-97-6764-1993,
bastos-J.Mol.Model.-20-2094-2014,
Burakov-Ecotox.Environ.Safe.-148-702-2018,
Baby-NanoscaleRes.Lett.-14-341-2019,
Cho-Sci.Rep.-10-7416-2020,
Maselugbo-J.Mater.Res.-37-4438-2022}. In addition, it is
reported that the adsorbed metal ions can be easily desorbed, and the
nanomaterials can be recycled and
reused~\cite{Baby-NanoscaleRes.Lett.-14-341-2019}. Furthermore, carbon-based
nanomaterials are highly biocompatible with living organisms and the
environment~\cite{Dutta-Inorganics-10-169-2022}.

When carbon atoms are replaced by boron
and nitrogen atoms, boron nitride nanotubes (BNNTs) are
obtained~\cite{Li-Phys.ELow-Dimens.Syst.Nanostructures-85-137-2017}. With an
identical structure to carbon nanotubes, BNNTs have similar mechanical
properties~\cite{Yin-Small-12-2942-2016,
Li-Phys.ELow-Dimens.Syst.Nanostructures-85-137-2017} and they are
electrically insulating, with a forbidden band between 5.0--6.0~eV, independent
of
chirality~\cite{Vaccarini-Carbon-38-1681-2000,
Li-Phys.ELow-Dimens.Syst.Nanostructures-85-137-2017}. In addition, the thermal
properties
are improved due to boron nitride (BN) having high atmospheric stability,
especially at high temperatures~\cite{ansari-Europ.J.Mech.ASolids-62-67-2017}.
This makes BNNT an excellent candidate
for applications in
electronics~\cite{Patki-J.Mater.Sci.Mater.Electron.-30-3899-2019},
sensors~\cite{Oh-Compos.Sci.Technol.-172-153-2019}, hydrogen
storage~\cite{Muthu-Renew.Energ.-85-387-2016},
medicine~\cite{Jedrzejczak-Silicka-Nanomaterials-8-605-2018,
Maiti-Front.Pharmacol.-9-1401-2019}, and also water
purification~\cite{Cho-Sci.Rep.-10-7416-2020,
Gonzalez-Ortiz-Mater.TodayAdv.-8-100107-2020,
Madeira-Appl.Surf.Sci.-573-151547-2022,
Maselugbo-J.Mater.Res.-37-4438-2022,
Turhan-Nanotechnology-33-242001-2022,
Khalid-RSCAdvances-12-6592-2022} among others.

Due to these particularities, research on a new structure has been
investigated, the boron nitride nanobelts (BN-nanobelts), an inorganic analog
of cyclofenacene synthesized in
2017~\cite{Merner-Angew.Chem.Int.Ed.-48-5487-2009,
Barbosa-Comput.Theor.Chem.-1208-113571-2022}. Aromatic BN-nanobelts have
radially oriented p orbitals with photoluminescent properties and excellent UV
absorber, suggesting that this molecule can be used as a UV
detector~\cite{Barbosa-Comput.Theor.Chem.-1208-113571-2022,
Povie-Science-356-172-2017,
Cheung-Chem-5-838-2019}.
In addition, BN-nanobelts have high chemical stability, thermal stability with
positive vibrational frequencies and insulating character with an estimated gap
of around 5~eV~\cite{Xia-Angew.Chemie.-133-10399-2021,
Barbosa-Comput.Theor.Chem.-1208-113571-2022}.
Considering this, properties
originating from these different topologies can help solve current problems
such as pollution caused by undesirable chemical contaminants that cause
adverse effects on nature and living
organisms~\cite{Baby-NanoscaleRes.Lett.-14-341-2019,
Wu-Environ.Pollut.-246-608-2019}. Contamination by heavy
metal ions, mainly in water resources, is a serious environmental problem.

In this study, the interaction of Cadmium (Cd), Nickel (Ni), and Lead (Pb),
with boron--nitride and M\"obius-type boron--nitride nanobelts were
investigated using the semiempirical tight binding theory. Several methods were
used to characterize the systems: best interaction
region detection, geometry optimization, molecular dynamics, electronic property
calculations, and topology studies.

\section{Materials and Methods}
\label{Sec:Method}

In this work, three heavy metals (Cd, Ni, and Pb) were used, together with two
different types of boron-nitride (BN) nanostructures.

\begin{figure}[htpb]
\centering
\begin{tabular}{ccc}
\subfigure[1DL]{\includegraphics[width=3.5cm]{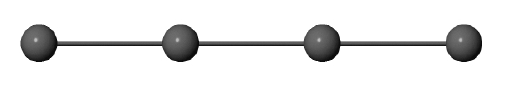} \label{Fig:1DL}}      &  \subfigure[1DZ]{\includegraphics[width=3.5cm]{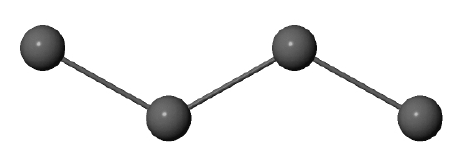} \label{Fig:1DZ}} \\
\subfigure[2D]{\includegraphics[width=3.5cm]{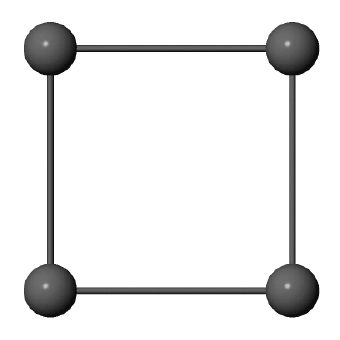} \label{Fig:2D}}        &  \subfigure[3D]{\includegraphics[width=3.5cm]{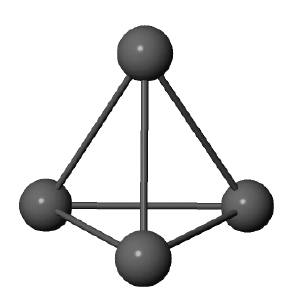} \label{Fig:3D}}
\end{tabular}
\caption{\label{Fig:Nanoclusters} Nanoclusters geometry: (a) one dimensional
linear chain (1DL),
(b) one dimensional zigzag chain (1DZ), (c) two dimensional plane (2D), and (d)
three dimensional tetrahedron (3D).}
\end{figure}

The metals considered here are in the form of four--atom nano\-clus\-ters (M4).
We considered a one-dimensional linear chain (1DL, Figure~\ref{Fig:1DL}), a
one-dimensional zigzag chain (1DZ, Figure~\ref{Fig:1DZ}), a two-dimensional
plane (2D, ~\ref{Fig:2D}), and a three-dimensional tetrahedron
(3D,~\ref{Fig:3D}).

\begin{figure}[htpb]
\centering
\begin{tabular}{cc}
\subfigure[BNNB]{\includegraphics[width=4.1cm]{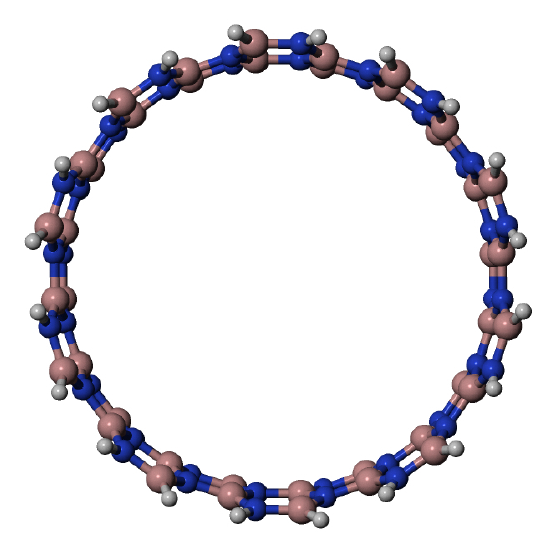}
\label{Fig:BNNB}}      &
\subfigure[MBNNB]{\includegraphics[width=3.5cm]{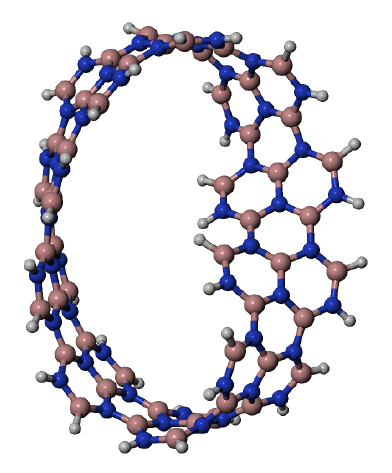}
\label{Fig:MBNNB}}
\end{tabular}
\caption{\label{Fig:Nanobelts} (a) Boron--nitride nanobelt (BNNB). (b) M\"obius
boron--nitride nanobelt (MBNNB).}
\end{figure}

Two types of BN nanostructures were considered: one consisting of a nanobelt
(NB) and the other consisting of a M\"obius nanobelt (twisted nanobelt). The
structures were generated using the Virtual NanoLab Atomistix Toolkit
software~\cite{VNL} as follows. We started with 2 unit cells of (10,0)
boron-nitride nanosheet repeated 10 times in the z direction and then wrapped
360 degrees. After that, the periodicity was removed, and the border atoms were
passivated with hydrogen. In the case of M\"obius nanobelts, after the initial
repetition of the cells, the nanobelt was twisted 180 degrees and then wrapped.
Both nanostructures are shown in Figure~\ref{Fig:Nanobelts}.

In order to clearly identify the systems, the following nomenclature was used:
BNNB for boron-nitride nanobelt, MBNNB for M\"obius boron-nitride nanobelt, and
BNNB+M4 (MBNNB+M4) for the complexes formed by the BNNB (MBNNB) and the metal
cluster, M4.

To perform the calculations, we used the semiempirical tight binding method, as
implemented in the xTB program~\cite{xTB_1,xTB_2}. The calculations were done
following a series of steps described below.

First, we optimized the structures of each individual system (two nanobelts and
four nanoclusters for each metal). Then, using the automated Interaction Site
Screening (aISS)~\cite{xTB-dock}, we generated different intermolecular
geometries (e.g., BNNB+M4 and MBNNB+M4) and subjected them to further genetic
optimization. We used the interaction energy (xTB-IFF)~\cite{xTB-IFF} of each
newly generated structure for ranking, and the genetic step was repeated ten
times until the best complex was obtained. The best--ranked complexes were
further subjected to structural optimization. In order to study the complexes'
stability, each structure obtained in the aISS step was subjected to a
molecular dynamic (MD) simulation for a period of 100~ps.

All geometry optimizations were performed using the GFN2--xTB method, which is
an accurate self--consistent method that includes multipole electrostatics and
density--dependent dispersion contributions~\cite{xTB_GFN2}. Extreme
optimization level was ensured, with a convergence energy of
$5\times10^{-8}$~E\textsubscript{h} and gradient norm
convergence of $5\times10^{-5}$~E\textsubscript{h}/a\textsubscript{0} (where
a\textsubscript{0} is the Bohr radius). MD
simulations were also conducted using the GFN2--xTB method.

The electronic properties calculated included the system energy, the energy of
the highest occupied molecular orbital (HOMO) ($\varepsilon_H$), the energy of
the lowest unoccupied molecular orbital (LUMO) ($\varepsilon_L$), the energy
gap between HOMO and LUMO orbitals ($\Delta \varepsilon =
\varepsilon_H - \varepsilon_L$, and the atomic charges using the CM5
scheme~\cite{charges_CM5}.

From the calculated charges, we estimated the charge transfer between the
isolated metal nanocluster and the nanobelts using the expression

\begin{equation}
\label{Eq:Qtransf}
\Delta {Q_{M4}} = Q_{M4}^{ads} - Q_{M4}^{iso},
\end{equation}
where $Q_{M4}^{ads}$ is the total charge on the metal nanocluster after
adsorption, and $Q_{M4}^{iso}$ is the total charge for the isolated metal
nanocluster.

The binding energies ($E_b$) of the adsorbed metals on the nanobelts were
calculated using the following expression

\begin{equation}
\label{Eq:bind}
E_b = E_{NB+M4} - E_{NB}- E_{M4}.
\end{equation}

In equation~\ref{Eq:bind}, $E_{NB}$ and $E_{M4}$ are the energies
for the isolated nanobelts and metal nanoclusters, respectively, and
$E_{NB+M4}$ is the energy of the NB+M4 complex (BNNB+M4 and MBNNB+M4 systems).

Finally, the topological characterization was carried out using the
MULTIWFN~\cite{multiwfn} software,where the topological properties such as
critical points, critical path, basins, etc. were determined for each complex.

\section{Results and discussion}
\label{Sec:results}
\subsection{Nanoclusters adsorption at the BN nanobelts}
\label{Sec:Geometry}

Figure~\ref{Fig:OPT_BNNB} shows the fully relaxed structures, with the point of
view being the
same as in Figure~\ref{Fig:Nanobelts}. Only the complexes with the lowest final
energy for each
metal are displayed. For BNNB, the Cd1DL, Ni1DZ, and Pb2D nanoclusters had the
lowest energy complexes. The MBNNB behaved similarly for cadmium and lead, but
for nickel, the optimal structure was with the Ni2D cluster. At first sight,
the interaction with nickel nanoclusters deforms the nanobelts to a greater
extent than the other metals and also forms more bonds. Both behaviors may
indicate greater binding between BNNB and MBNNB with Ni.

\renewcommand{\sizeA}{3.0cm}
\begin{figure}[htpb]
\centering
\begin{tabular}{ccc}
\subfigure[BNNB+Cd1DL]{\includegraphics[width=\sizeA]{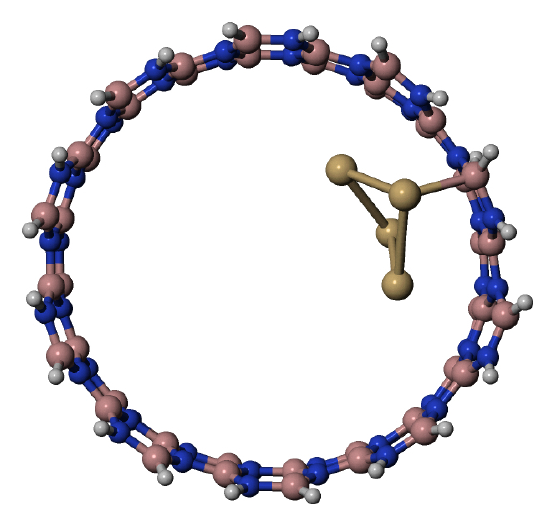}
\label{subFig:BNNB+Cd1DL}}
      &
\subfigure[BNNB+Ni1DZ]{\includegraphics[width=\sizeA]{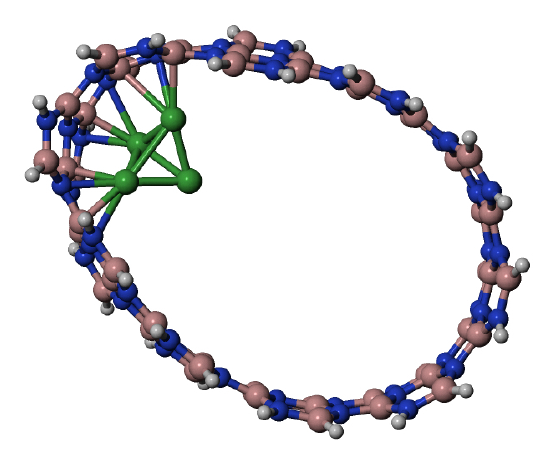}
\label{subFig:BNNB+Ni1DZ}}
      &
\subfigure[BNNB+Pb2D]{\includegraphics[width=\sizeA]{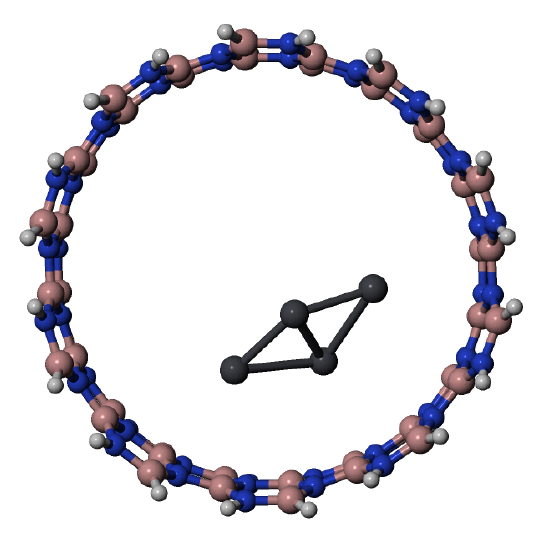}
\label{subFig:BNNB+Pb2D}} \\

\subfigure[MBNNB+Cd1DL]{\includegraphics[width=\sizeA]{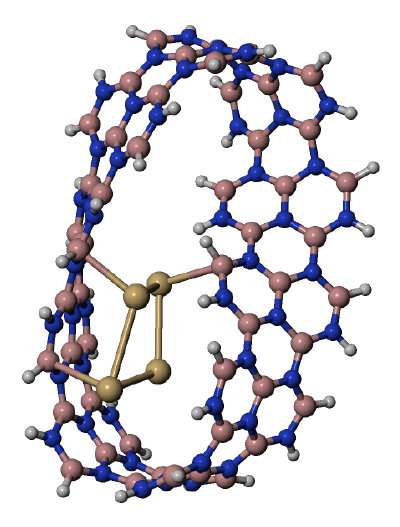}
\label{subFig:MBNNB+Cd1DL}}
      &
\subfigure[MBNNB+Ni2D]{\includegraphics[width=\sizeA]{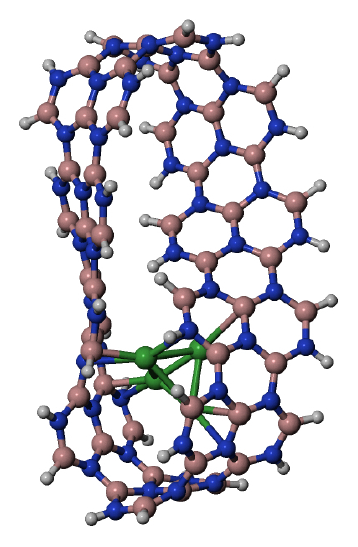}
\label{subFig:MBNNB+Ni2D}}
      &
\subfigure[MBNNB+Pb2D]{\includegraphics[width=\sizeA]{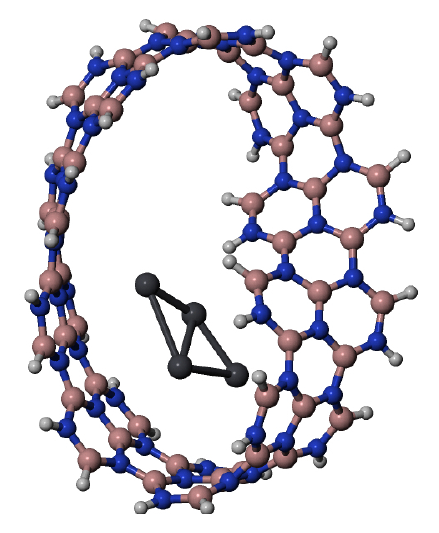}
\label{subFig:MBNNB+Pb2D}}

\end{tabular}
\caption{\label{Fig:OPT_BNNB}  Fully relaxed complexes with lowest energies.
Image rendered with Jmol software~\cite{Jmol}.}
\end{figure}

A significant contrast between chemisorption and physisorption is the effect on
the electronic states of both the adsorbate and the
adsorbent~\cite{Everett2001}. We will observe in sections ~\ref{Sec:ElecProp}
and~\ref{Sec:Topo} that the electronic properties of the systems are altered
upon adsorption, confirming this phenomenon.

\begin{table}[htpb]
\caption{Data from geometry optimization and electronic calculation$^\dagger$.}
\label{Tab:DataResults}
\begin{center}
\begin{tabular}{lrrrrrr}
  \hline
  Complex & $E_{b}^{aISS}$ & $E_b$ & $\Delta Q_M$ & Gap ($\Delta \varepsilon$)
  & Distances \\
  \hline
  \hline
  BNNB+Cd1DL    & -95.33    &  -82.87  & 0.0622   &  0.999 &
2.48/3.03/3.18/3.23/3.32/3.54 \\

  BNNB+Ni1DZ    & -142.77   &  -116.01 & -0.1195  &  0.144 &
  2.24/2.27/2.32/2.44 \\
  BNNB+Pb2D     & -93.36    &  -54.31  & 0.0014  &  1.226 & 3.51/3.53/3.71 \\
  \hline
  MBNNB+Cd1DL   & -109.25   &  -96.78  & 0.0085   &  0.826 &
  2.59/2.68/2.85/2.88/2.90/3.38 \\
                &           &          &          &         &
                3.39/3.41/3.58/3.69/3.83/3.92 \\
  MBNNB+Ni2D    & -163.68   &  -149.83 & -0.2223  &  0.109 &
  2.15/2.21/2.52/2.57/2.70/2.71/3.60 \\
  MBNNB+Pb2D    & -99.00    &  -59.98  & 0.0248  &  1.449 &
  2.81/2.92/3.31/3.37/3.61/4.41
 \\
  \hline
\end{tabular}
\begin{flushleft}
\tiny {$^\dagger$ $E_{b}^{aISS}$ and $E_{b}$ are in units kcal/mol, $\Delta Q_M$
are in units of $e$, $\Delta \varepsilon$ is in units of $eV$ and the distances
are in \AA, respectively.}
\end{flushleft}
\end{center}
\end{table}

Table~\ref{Tab:DataResults} displays the data for the complexes with the lowest
energy. Two binding energies are reported: $E_{b}^{aISS}$ and $E_b$. The former
is obtained from the aISS step, while the latter is calculated using
equation~\ref{Eq:bind} for the fully optimized individual structures and final
complex. Comparing both binding energies for each system, the Ni clusters have
the lowest binding energies, indicating stronger adsorption, followed by Cd and
Pb. As all the binding energies are negative, all the adsorption processes are
favorable.

The graphical representation of the complexes in Figure~\ref{Fig:OPT_BNNB}
suggests that only Cd and Ni formed bonds with the nanobelts. In the case of
Cd, it made one bond with BNNB and two bonds with MBNNB. The Ni cluster formed
ten bonds with BNNB and eight with MBNND, whereas Pb did not make any bonds
with either BNNB or MBNNB. Since the bond information from
Figure~\ref{Fig:OPT_BNNB} is only based on geometrical data, the Quantum Theory
of Atoms in Molecule (QTAIM)~\cite{bader1994} was used in
Section~\ref{Sec:Topo} as a more accurate method to study bond formation.

Geometry optimization involves using an algorithm to obtain a local minimum
structure on the potential energy surface (PES), which allows us to determine
the lowest energy conformers of a system. However, this method provides no
information about the system's stability over time. Molecular dynamics
simulation, on the other hand, analyzes the movement of atoms and molecules at
a specific temperature (here, 298.15~K) and provides a way to explore the PES.
We used this method to perform simulations on each complex, starting with the
structures obtained from the aISS step as initial conformations. The
simulations were run for a production time frame of 100~ps with a time step of
2~fs and an optional dump step of 50~fs, at which the final structure was
written to a trajectory file.

Figures~\ref{Fig:MD_BNNB} and~\ref{Fig:MD_MBNNB} show the system frames at
several simulation times (0 ps, 25 ps, 50 ps, 75 ps, and 100 ps). In all cases,
the metal nanoclusters remained bound to their respective BN nanobelts,
indicating system stability. By comparing the snapshots from
figures~\ref{Fig:MD_BNNB}
and~\ref{Fig:MD_MBNNB} with Figure~\ref{Fig:Nanobelts}, we confirmed that the
Ni nanocluster caused the most significant modifications to the nanobelt. The
full molecular dynamics movies can be downloaded from the Zenodo
server~\cite{aguiar_c_2023_7662326}.

\renewcommand{\sizeA}{3.0cm}
\begin{figure}[htpb]
\centering
\begin{tabular}{ccccc}
\subfigure[0~ps]{\includegraphics[width=\sizeA]{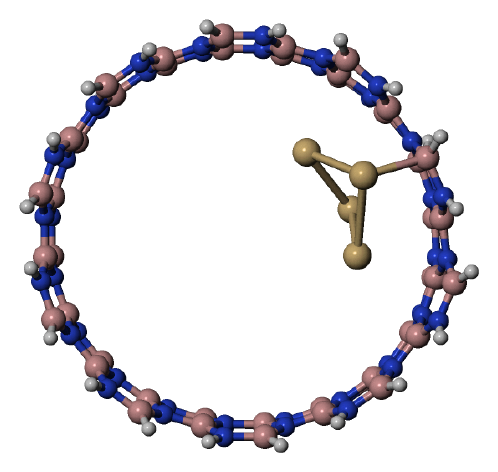}}
      &
\subfigure[25~ps]{\includegraphics[width=\sizeA]{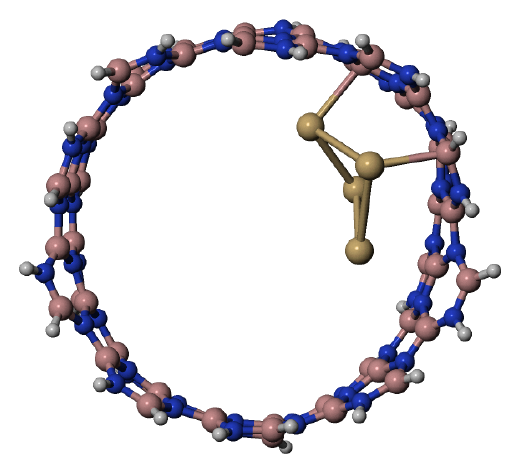}}
      &
\subfigure[50~ps]{\includegraphics[width=\sizeA]{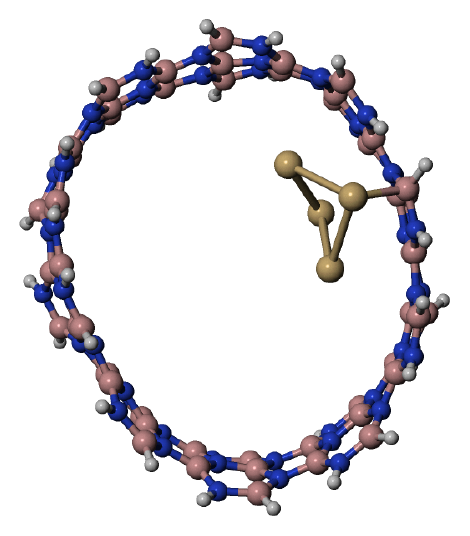}}
      &
\subfigure[75~ps]{\includegraphics[width=\sizeA]{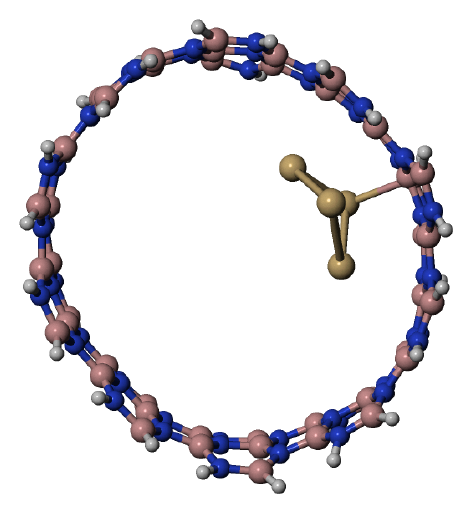}}
      &
\subfigure[100~ps]{\includegraphics[width=\sizeA]{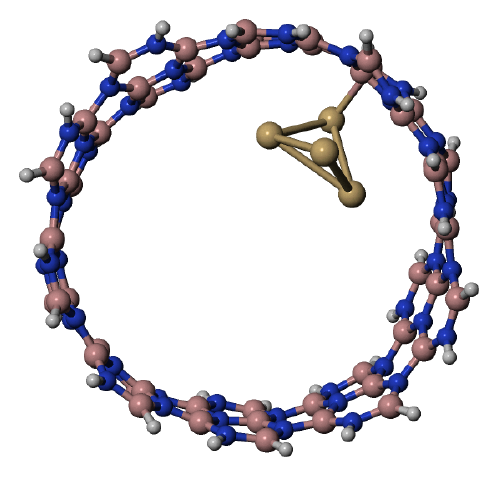}
} \\
\subfigure[0~ps]{\includegraphics[width=\sizeA]{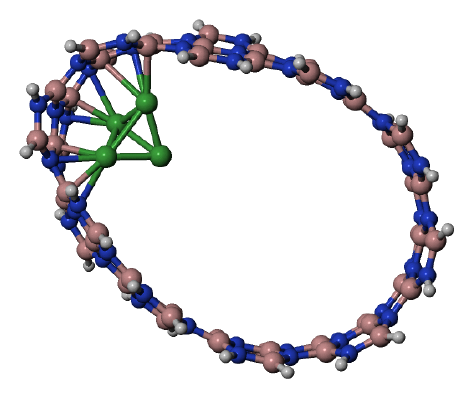}}
     &
\subfigure[25~ps]{\includegraphics[width=\sizeA]{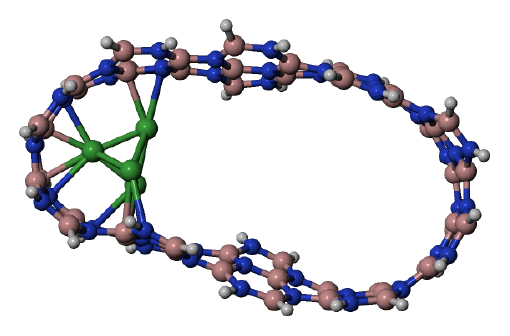}}
      &
\subfigure[50~ps]{\includegraphics[width=\sizeA]{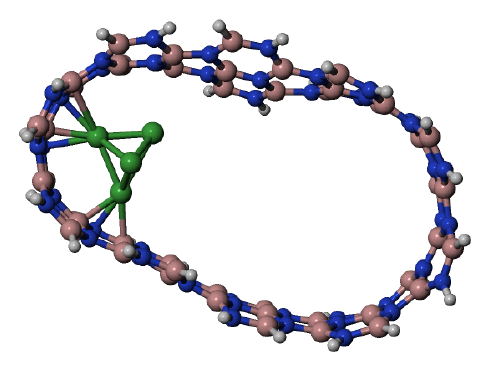}}
      &
\subfigure[75~ps]{\includegraphics[width=\sizeA]{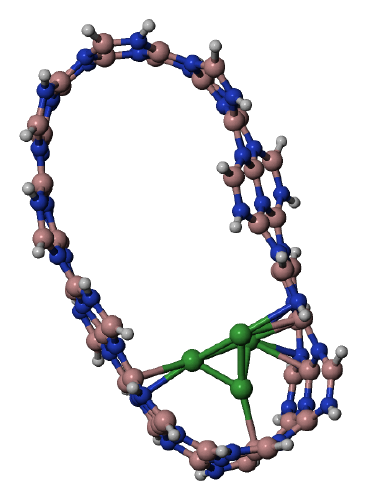}}
      &
\subfigure[100~ps]{\includegraphics[width=\sizeA]{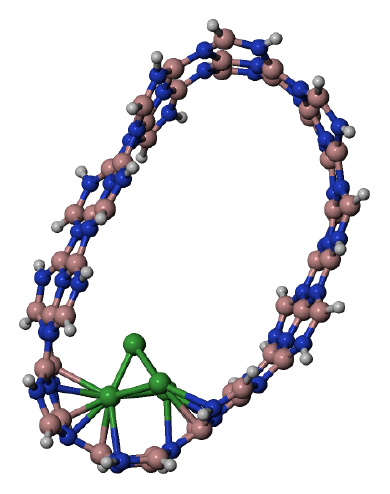}}
 \\
\subfigure[0~ps]{\includegraphics[width=\sizeA]{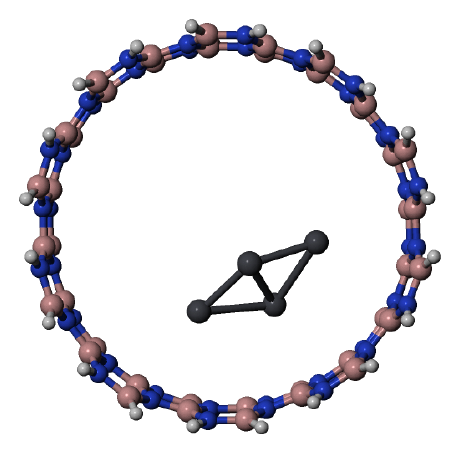}}
     &
\subfigure[25~ps]{\includegraphics[width=\sizeA]{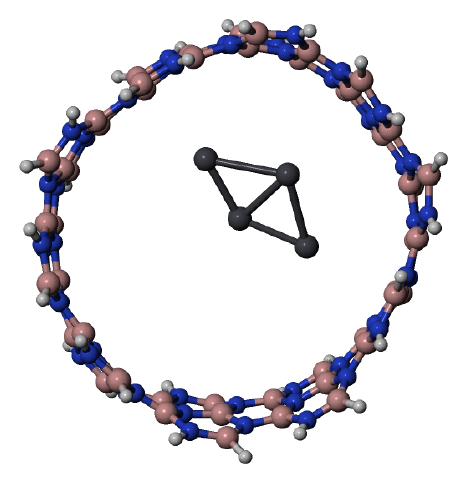}}
      &
\subfigure[50~ps]{\includegraphics[width=\sizeA]{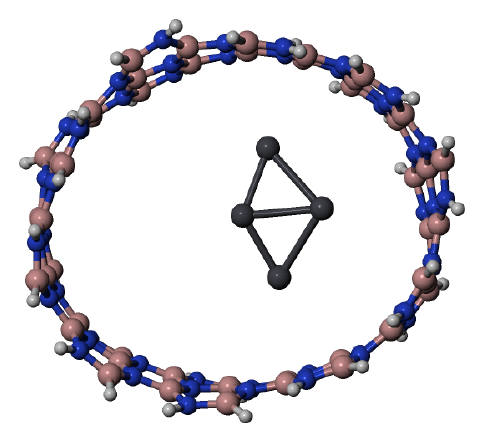}}
      &
\subfigure[75~ps]{\includegraphics[width=\sizeA]{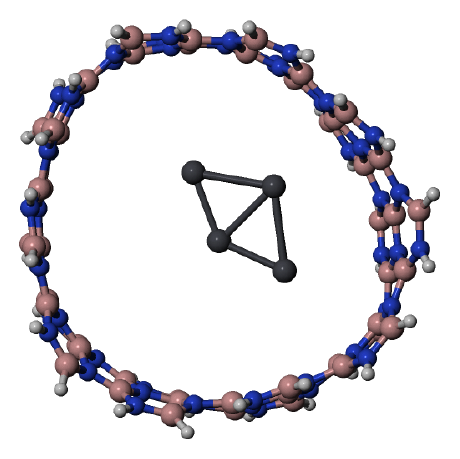}}
      &
\subfigure[100~ps]{\includegraphics[width=\sizeA]{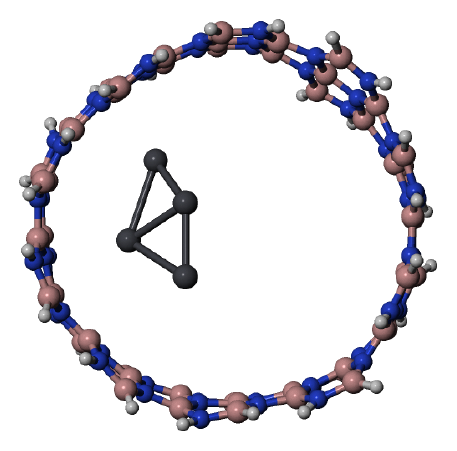}}

\end{tabular}
\caption{\label{Fig:MD_BNNB}  Molecular dynamics snapshots at different
simulation times for BNNB complexes. Simulation
times (from left to right): 0~ps, 25~ps, 50~ps, 75~ps and 100~ps, respectively.
And
systems (from top to bottom): BNNB+Cd1DL, BNNB+Ni1DZ and BNNB+Pb2D,
respectively.}
\end{figure}

\renewcommand{\sizeA}{3.0cm}
\begin{figure}[tbph]
\centering
\begin{tabular}{ccccc}
\subfigure[0~ps]{\includegraphics[width=\sizeA]{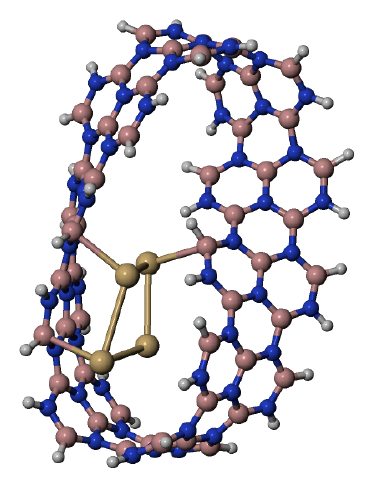}}
      &
\subfigure[25~ps]{\includegraphics[width=\sizeA]{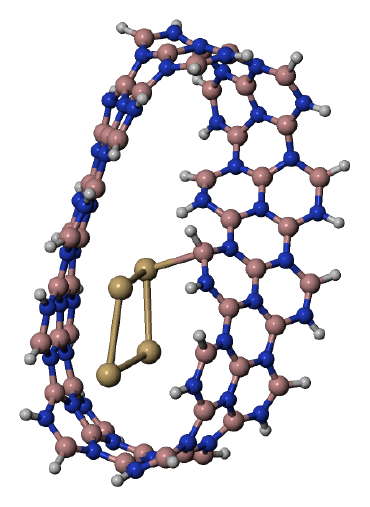}}
      &
\subfigure[50~ps]{\includegraphics[width=\sizeA]{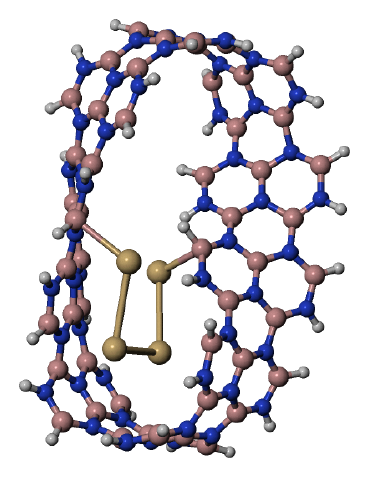}}
      &
\subfigure[75~ps]{\includegraphics[width=\sizeA]{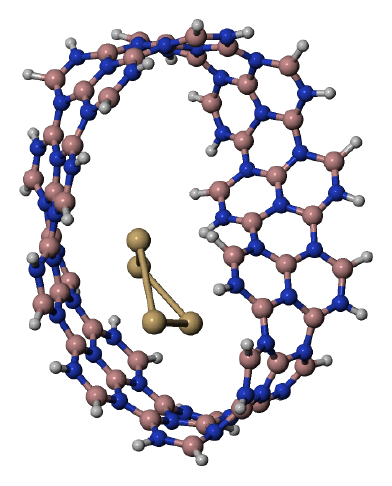}}
      &
\subfigure[100~ps]{\includegraphics[width=\sizeA]{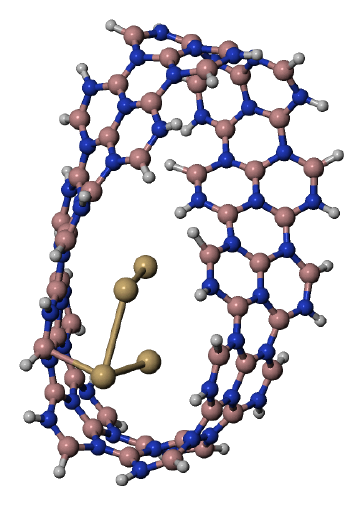}
} \\
\subfigure[0~ps]{\includegraphics[width=\sizeA]{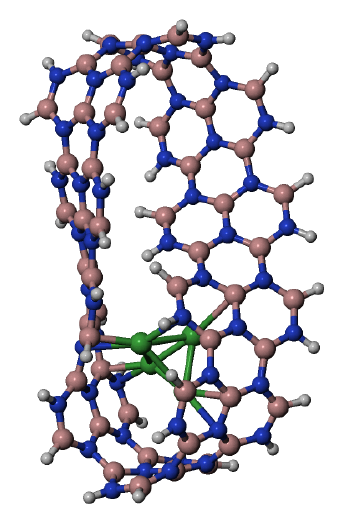}}
     &
\subfigure[25~ps]{\includegraphics[width=\sizeA]{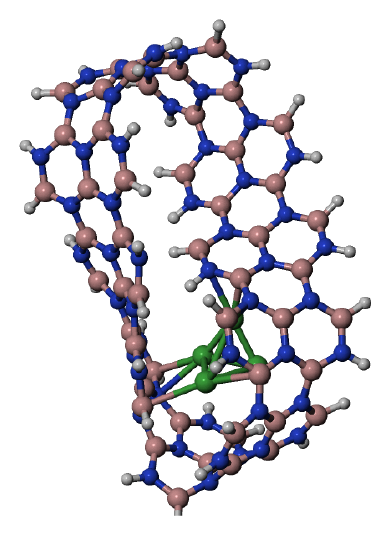}}
      &
\subfigure[50~ps]{\includegraphics[width=\sizeA]{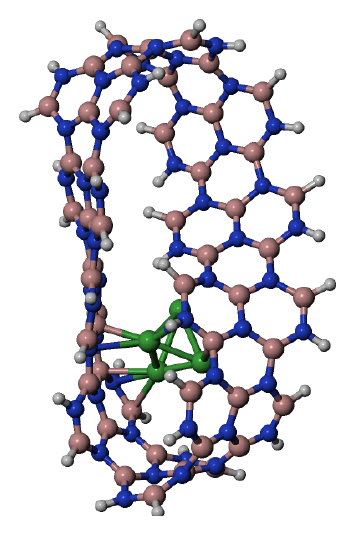}}
      &
\subfigure[75~ps]{\includegraphics[width=\sizeA]{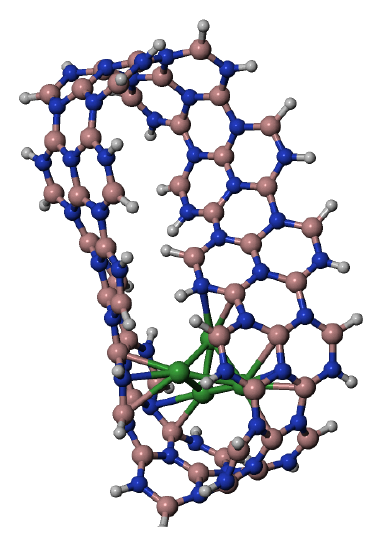}}
      &
\subfigure[100~ps]{\includegraphics[width=\sizeA]{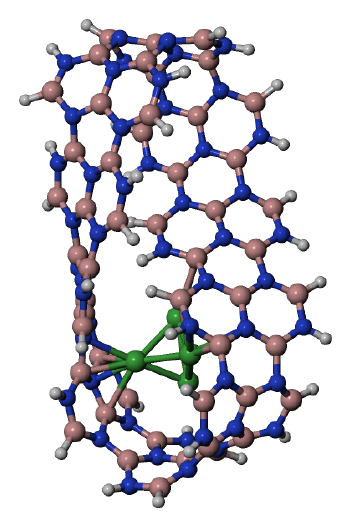}}
 \\
\subfigure[0~ps]{\includegraphics[width=\sizeA]{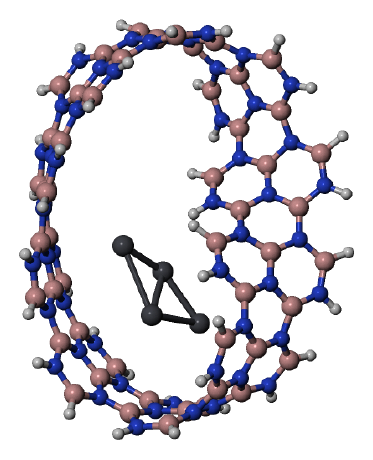}}
     &
\subfigure[25~ps]{\includegraphics[width=\sizeA]{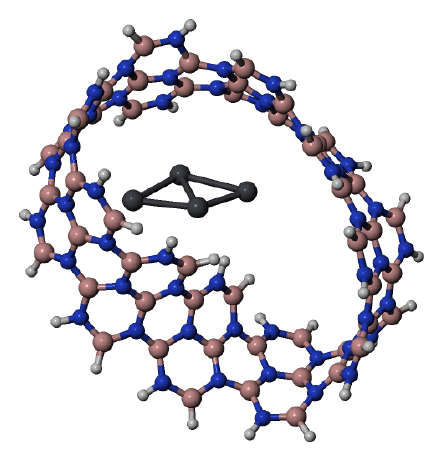}}
      &
\subfigure[50~ps]{\includegraphics[width=\sizeA]{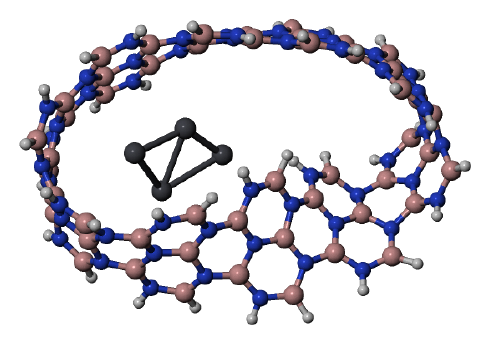}}
      &
\subfigure[75~ps]{\includegraphics[width=\sizeA]{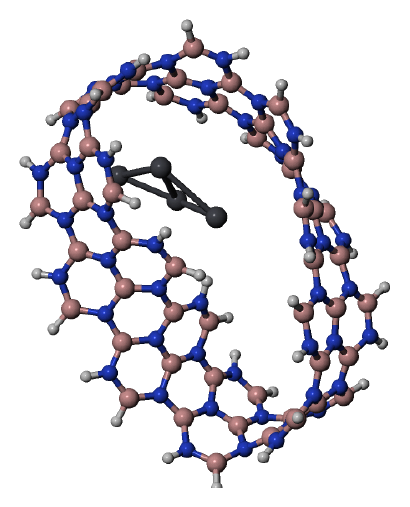}}
      &
\subfigure[100~ps]{\includegraphics[width=\sizeA]{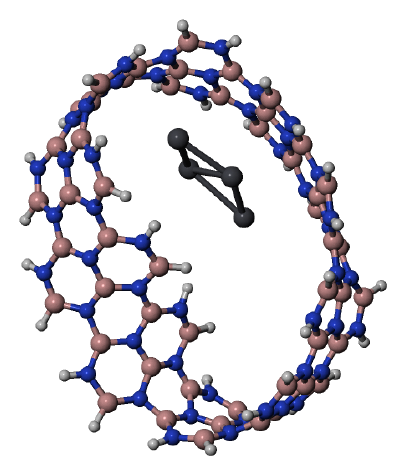}}

\end{tabular}
\caption{\label{Fig:MD_MBNNB}  Molecular dynamics snapshots at different
simulation times for MBNNB complexes. Simulation
times (from left to right): 0~ps, 25~ps, 50~ps, 75~ps and 100~ps, respectively.
And
systems (from top to bottom): MBNNB+Cd1DL, MBNNB+Ni2D and MBNNB+Pb2D,
respectively.}
\end{figure}

\subsection{Electronic properties}
\label{Sec:ElecProp}

Figures~\ref{Fig:HOMO} and~\ref{Fig:LUMO} show the calculated highest
occupied/lowest unoccupied molecular orbitals (HOMO/LUMO) for all optimized
systems.

The molecular orbitals for the BNNB system (Figures~\ref{Fig:HOMO_BNNB}
and~\ref{Fig:LUMO_BNNB}) are homogeneously distributed over the entire
structure, as expected due to the belt symmetry. These distributions change
when the nanobelts are twisted, breaking their symmetry
(figures~\ref{Fig:HOMO_MBNNB} and~\ref{Fig:LUMO_MBNNB}). In the M\"obius
nanobelt, there is a redistribution of both molecular orbital surfaces, with
more volume concentrated around the regions with the twist, especially for the
HOMO. The electronic gap for the BNNB is equal to 3.973~eV, which is slightly
decreased to 3.859~eV for the MBNNB.

In the case of the complexes, the interaction with the heavy metal nanoclusters
greatly modifies the HOMO for all systems, with the orbital volume becoming
more concentrated around the metal adsorption regions. A similar behavior
occurs for the LUMO, but to a greater extent for Ni and Pb than for Cd. The gap
for the complexes is also modified after adsorption, with the greatest change
observed for Ni, followed by Cd and Pb. The interaction of both nanobelts with
all the metals decreases the gap, as shown in Table~\ref{Tab:DataResults}.
Again, the interaction with Ni induced the greatest change in the electronic
gap.

Another indicator of adsorption strength is the charge transfer ($\Delta
{Q_{M4}}$) between the adsorbent and adsorbate. The $\Delta {Q_{M4}}$ was
calculated using the equation~\ref{Eq:Qtransf} and its value is shown in
Table~\ref{Tab:DataResults} for all systems. In both cases (BNNB and MBNNB),
the greatest charge transfer occurs between the nanobelts and the Ni
nanocluster.

Based on the intensity of the modifications in the electronic properties of the
BNNB and MBNNB after interaction with metals, we can conclude that
chemisorption occurs for all systems. This conclusion from the analysis of
electronic properties will be confirmed by the topological studies presented in
the next section.

\renewcommand{\sizeA}{4.0cm}
\begin{figure}[tbph]
\centering
\begin{tabular}{cccc}
\subfigure[BNNB]{\includegraphics[width=\sizeA]{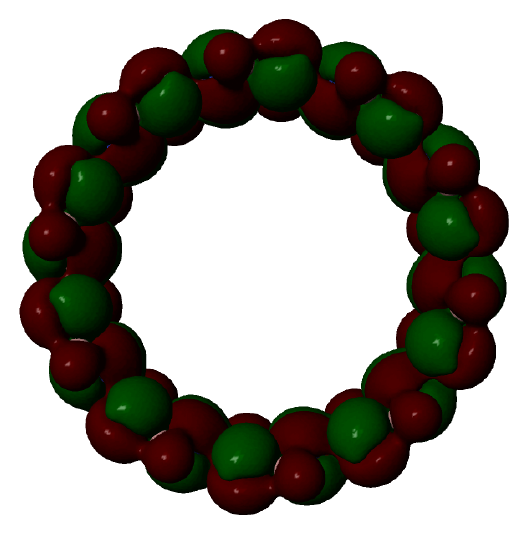}
\label{Fig:HOMO_BNNB}}      &
\subfigure[BNNB+Cd1DL]{\includegraphics[width=\sizeA]{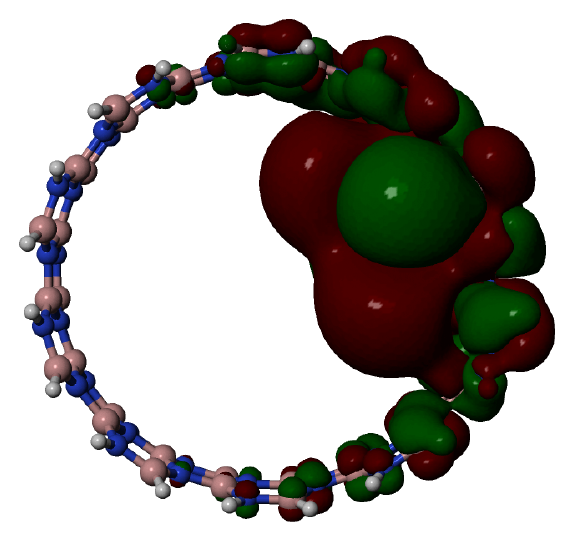}
\label{Fig:HOMO_BNNB+Cd1DL}}      &
\subfigure[BNNB+Ni1DZ]{\includegraphics[width=\sizeA]{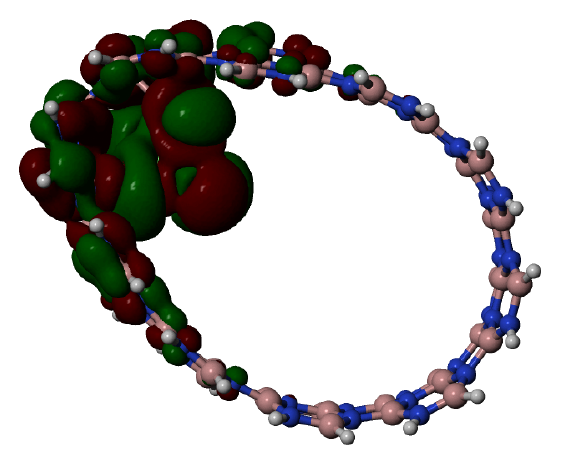}
\label{Fig:HOMO_BNNB+Ni1DZ}}      &
\subfigure[BNNB+Pb2D]{\includegraphics[width=\sizeA]{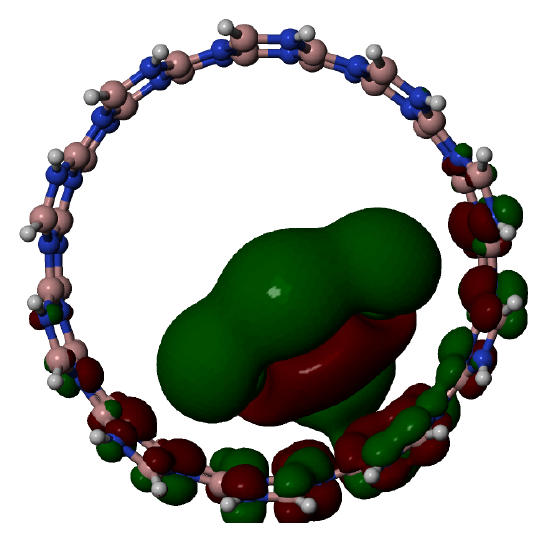}
\label{Fig:HOMO_BNNB+Pb2D}} \\

\subfigure[MBNNB]{\includegraphics[width=\sizeA]{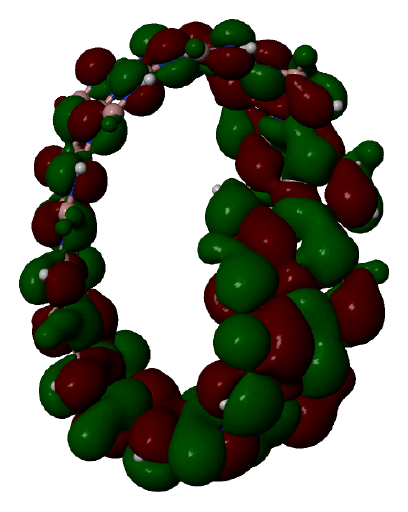}
\label{Fig:HOMO_MBNNB}}      &
\subfigure[MBNNB+Cd1DL]{\includegraphics[width=\sizeA]{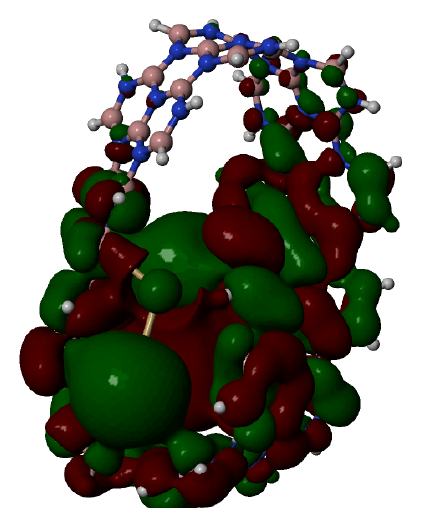}

\label{Fig:HOMO_MBNNB+Cd1DL}}      &
\subfigure[MBNNB+Ni2D]{\includegraphics[width=\sizeA]{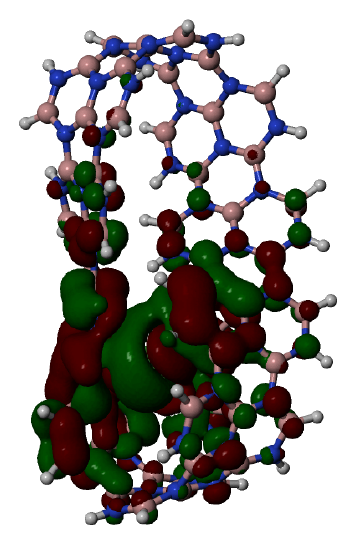}
\label{Fig:HOMO_MBNNB+Ni2D}}      &
\subfigure[MBNNB+Pb2D]{\includegraphics[width=\sizeA]{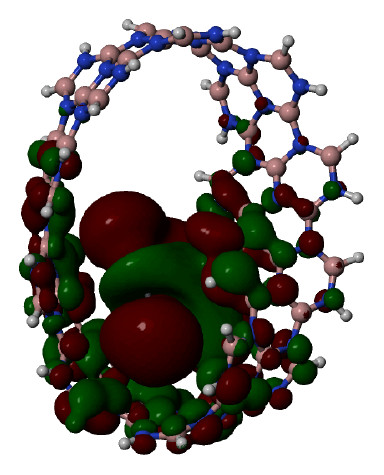}
\label{Fig:HOMO_MBNNB+Pb2D}} \\
\end{tabular}
\caption{\label{Fig:HOMO} Highest occupied molecular orbital (HOMO) for all
systems. Red (green) color
represents negative (positive) values. Orbital surfaces rendered with with
isovalue equal to 0.001 and with Jmol software~\cite{Jmol}.}
\end{figure}

\renewcommand{\sizeA}{4.0cm}
\begin{figure}[tbph]
\centering
\begin{tabular}{cccc}
\subfigure[BNNB]{\includegraphics[width=\sizeA]{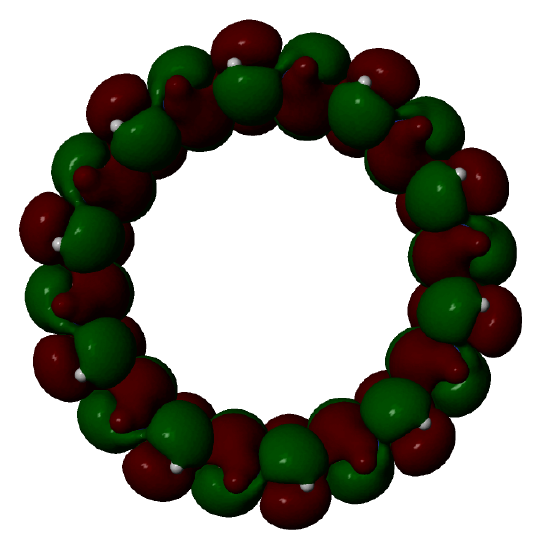}
\label{Fig:LUMO_BNNB}}      &
\subfigure[BNNB+Cd1DL]{\includegraphics[width=\sizeA]{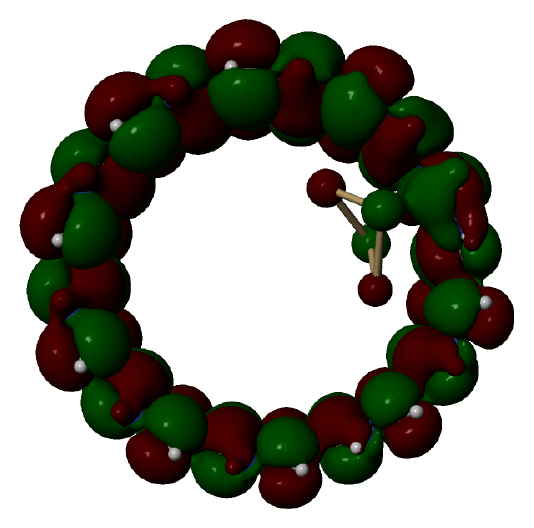}
\label{Fig:LUMO_BNNB+Cd1DL}}      &
\subfigure[BNNB+Ni1DZ]{\includegraphics[width=\sizeA]{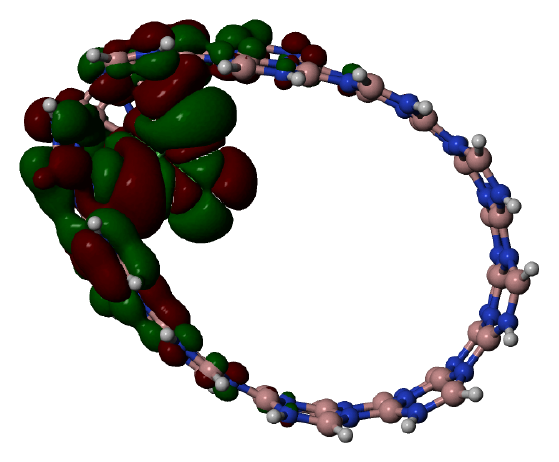}
\label{Fig:LUMO_BNNB+Ni1DZ}}      &
\subfigure[BNNB+Pb2D]{\includegraphics[width=\sizeA]{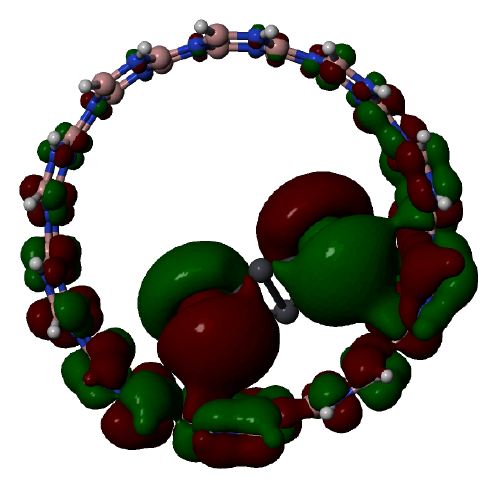}
\label{Fig:LUMO_BNNB+Pb2D}} \\
\subfigure[MBNNB]{\includegraphics[width=\sizeA]{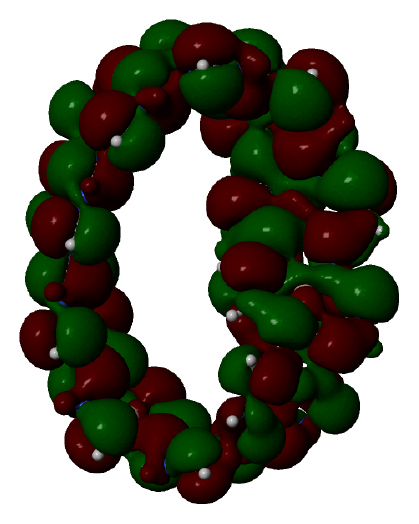}
\label{Fig:LUMO_MBNNB}}      &
\subfigure[MBNNB+Cd1DL]{\includegraphics[width=\sizeA]{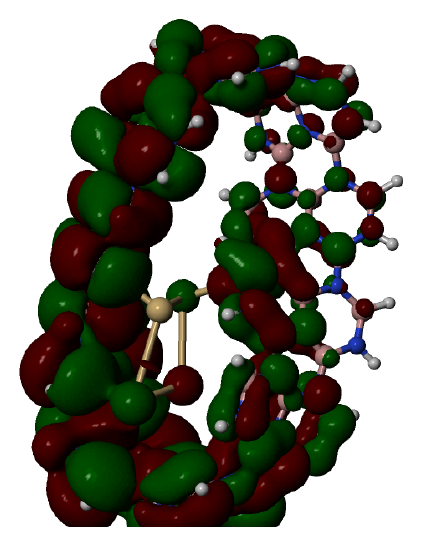}
\label{Fig:LUMO_MBNNB+Cd1DL}}      &
\subfigure[MBNNB+Ni2D]{\includegraphics[width=\sizeA]{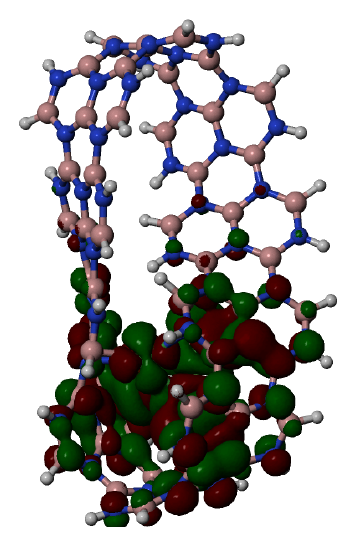}
\label{Fig:LUMO_MBNNB+Ni2D}}      &
\subfigure[MBNNB+Pb2D]{\includegraphics[width=\sizeA]{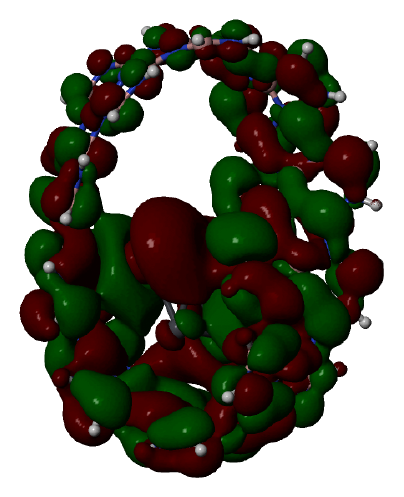}
\label{Fig:LUMO_MBNNB+Pb2D}} \\
\end{tabular}
\caption{\label{Fig:LUMO} Lowest unoccupied molecular orbital (LUMO) for all
systems. Red (green) color
represents negative (positive) values. Orbital surfaces rendered with with
isovalue equal to 0.001 and with Jmol software~\cite{Jmol}.}
\end{figure}

\subsection{Topological analysis}
\label{Sec:Topo}

In topological analysis, a key objective is to identify critical points. These
critical points (CPs) are points where the gradient norm of the function value
is zero, and they are classified into four types based on the negative
eigenvalues of the Hessian matrix of the real function~\cite{bader1994}.

When three eigenvalues of the Hessian matrix are negative, the CPs are known as
nuclear critical points (NCPs) and their positions are almost identical to
atomic positions. They are represented as $\bf (3,-3)$. If two eigenvalues of
the Hessian matrix are negative, the CPs are represented as $\bf (3,-1)$ and
are referred to as bond critical points (BCPs). These BCPs generally appear
between atom pairs in electron density analysis and are related to the strength
and type of bonds formed. The value of the electron density ($\rho$) and the
sign of its Laplacian ($\nabla ^2\rho$) at the  $\bf (3,-1)$ CPs can be related
to the strength and type of the bonds  ~\cite{Matta2007}. CPs with one negative
eigenvalue are represented as $\bf (3,+1)$ and are called ring critical points
(RCPs) because they usually appear in the center of a ring system. Finally, CPs
with no negative eigenvalues are represented as $\bf (3,+3)$ and are called
cage critical points (CCPs) because they typically appear in the center of a
cage system during electron density analysis.

Diagrams with the 3D distribution of critical points for each complex are shown
in Figure~\ref{Fig:3D-CPs}. The orange dots represent the bond critical points
(BCPs), the yellow dots represent the ring critical points (RCPs), and the
green dots represent the cage critical points. All the complexes show several
critical points, indicating a favorable interaction between the metals and the
belts. In all cases, the number of critical points made between the metal
nanocluster and the M\"obious nanobelts is greater than with the nanobelts
alone.
This can be associated with the fact that M\"obious belts form like small
pockets
where the metal nanocluster can be docked. The interaction between the Ni2D
nanocluster with the MBNNB is so intense that it pulled both sides of the belt
in such a way that new bonds are formed between both sides (see
Figure~\ref{Fig:topo_MBNNB+Ni2D}). For
better visualization of the formed critical points, movies with spinning
structures can be downloaded from the Zenodo
server~\cite{aguiar_c_2023_7662326}. Another confirmation
that the strongest interactions are between the Ni nanoclusters and the
boron-nitride nanobelts is the lowest bond distances shown in
Table~\ref{Tab:DataResults}.

\renewcommand{\sizeA}{4.0cm}
\begin{figure}[tbph]
\centering
\begin{tabular}{cccc}
\subfigure[BNNB+Cd1DL]{\includegraphics[width=\sizeA]{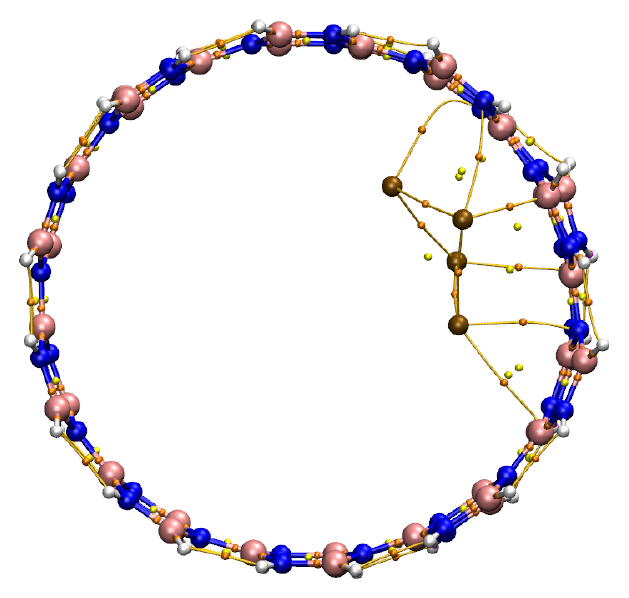}
\label{Fig:topo_BNNB+Cd1DL}}      &
\subfigure[BNNB+Ni1DZ]{\includegraphics[width=\sizeA]{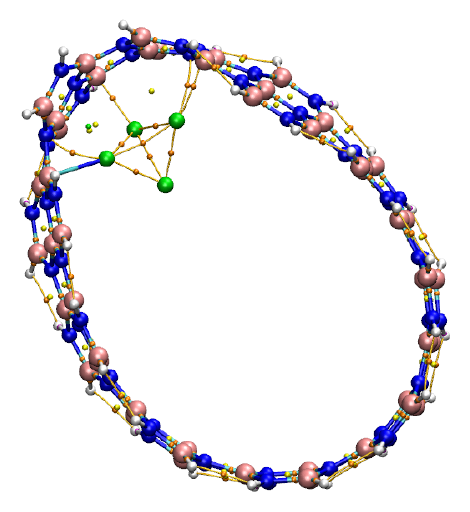}
\label{Fig:topo_BNNB+Ni1DZ}}      &
\subfigure[BNNB+Pb2D]{\includegraphics[width=\sizeA]{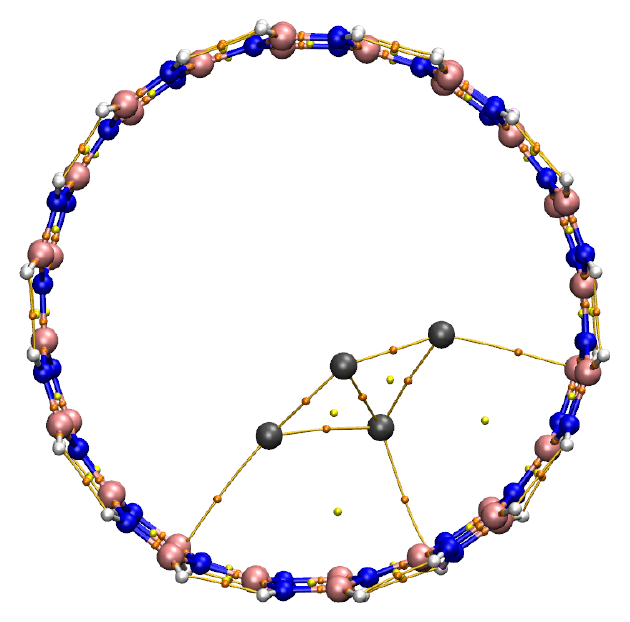}
\label{Fig:topo_BNNB+Pb2D}} \\
\subfigure[MBNNB+Cd1DL]{\includegraphics[width=\sizeA]{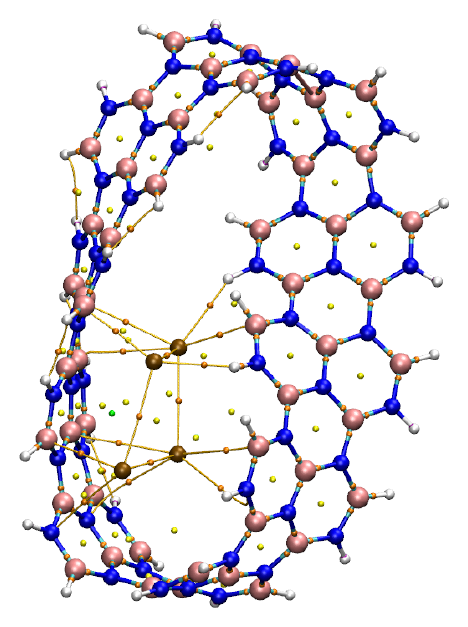}
\label{Fig:topo_MBNNB+Cd1DL}}      &
\subfigure[MBNNB+Ni2D]{\includegraphics[width=\sizeA]{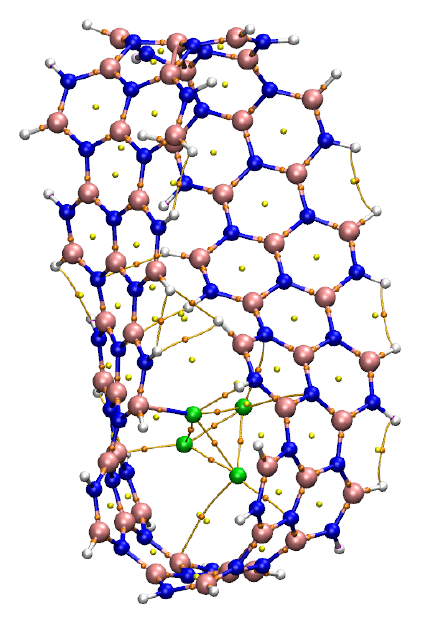}
\label{Fig:topo_MBNNB+Ni2D}}      &
\subfigure[MBNNB+Pb2D]{\includegraphics[width=\sizeA]{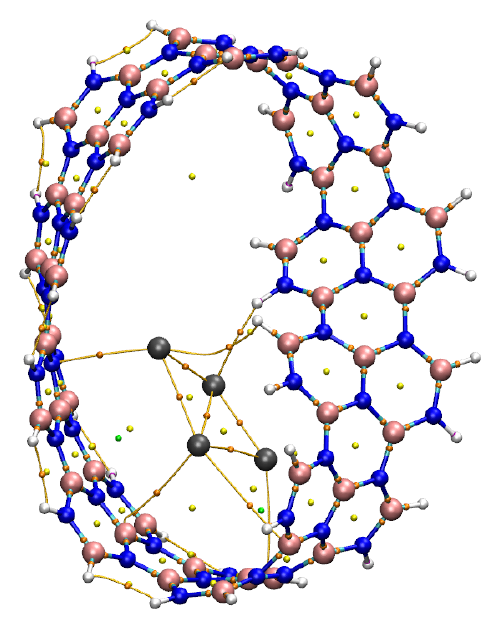}
\label{Fig:topo_MBNNB+Pb2D}} \\
\end{tabular}
\caption{\label{Fig:3D-CPs} Detected critical points: BCPs (orange dots), RCPs
(yellow dots), and CCPs (green dots). Image rendered with VMD
software~\cite{vmd}.}
\end{figure}

The strength of a bond, whether it is covalent or non--covalent, can be
classified by examining the electron density ($\rho$) and the sign of its
Laplacian ($\nabla^2\rho$). A value of $\rho$ greater than 0.20~a.u. indicates
a covalent bond, while a value less than 0.10~a.u. indicates a non--covalent
bond. Moreover, a bond can be classified as covalent if $\nabla^2\rho$ is less
than 0, and as non--covalent if $\nabla^2\rho$ is greater than
0~\cite{Matta2007}. The confinement of electron movement is related to the ELF
index, which takes values in the range of 0 to 1~\cite{elf,elf2}. Large values
of the ELF index indicate that electrons are highly localized, which is
indicative of the presence of a covalent bond. The LOL index is another
function that is used to identify regions of high localization~\cite{lol}.
Values of the LOL index are also in the range of 0 to 1, with smaller (larger)
values typically appearing in the boundary (inner) regions.

Figure~\ref{Fig:Topol} shows the electron density ($\rho$), Laplacian of the
electron density ($\nabla ^2\rho$), electron localization function (ELF) index,
and localized orbital locator (LOL) index at all the detected bond critical
points. The maximum number of BCPs for MBNNB is twice that for BNNB, indicating
that using M\"obius boron–nitride nanobelts to capture heavy metal nanoclusters
is a better choice. According to the classification criteria above, higher
values of $\rho$ indicate stronger bonds.
Figures~\ref{Fig:Rho_BNNB},~\ref{Fig:Lap_BNNB},~\ref{Fig:Rho_MBNNB},
and~\ref{Fig:Lap_MBNNB} show
that MBNNB made stronger bonds with Ni clusters (green regions) than with the
other metals (orange regions for Cd clusters and gray regions for Pb clusters).
Even though Cd nanocluster forms more bonds with MBNNB than Ni nanocluster,
they have lower values of $\rho$ and $\nabla^2\rho$. The ELF
(figures~\ref{Fig:ELF_BNNB}, and \ref{Fig:ELF_MBNNB}) and LOL
(figures~\ref{Fig:LOL_BNNB}, and \ref{Fig:LOL_MBNNB}) indexes show that some of
the BCPs for Cd nanoclusters
have greater values than for Ni and Pb clusters. This higher electron
localization of Cd cluster is associated with the HOMO redistribution shown in
figures~\ref{Fig:HOMO_BNNB+Cd1DL},
and~\ref{Fig:HOMO_MBNNB+Cd1DL}. The topological analysis confirms that both
boron–nitride nanobelts are capable of adsorbing the three heavy metal
nanoclusters studied here. Nevertheless, the MBNNB presented greater values for
all the descriptors used. In all cases, the Ni nanoclusters are chemisorbed,
whereas Cd and Pb nanoclusters are physisorbed.

\renewcommand{\sizeA}{3.5cm}
\begin{figure}[tbph]
\centering
\begin{tabular}{ccccc}
\subfigure[BNNB+M4 ($\rho$)]{\includegraphics[width=\sizeA]{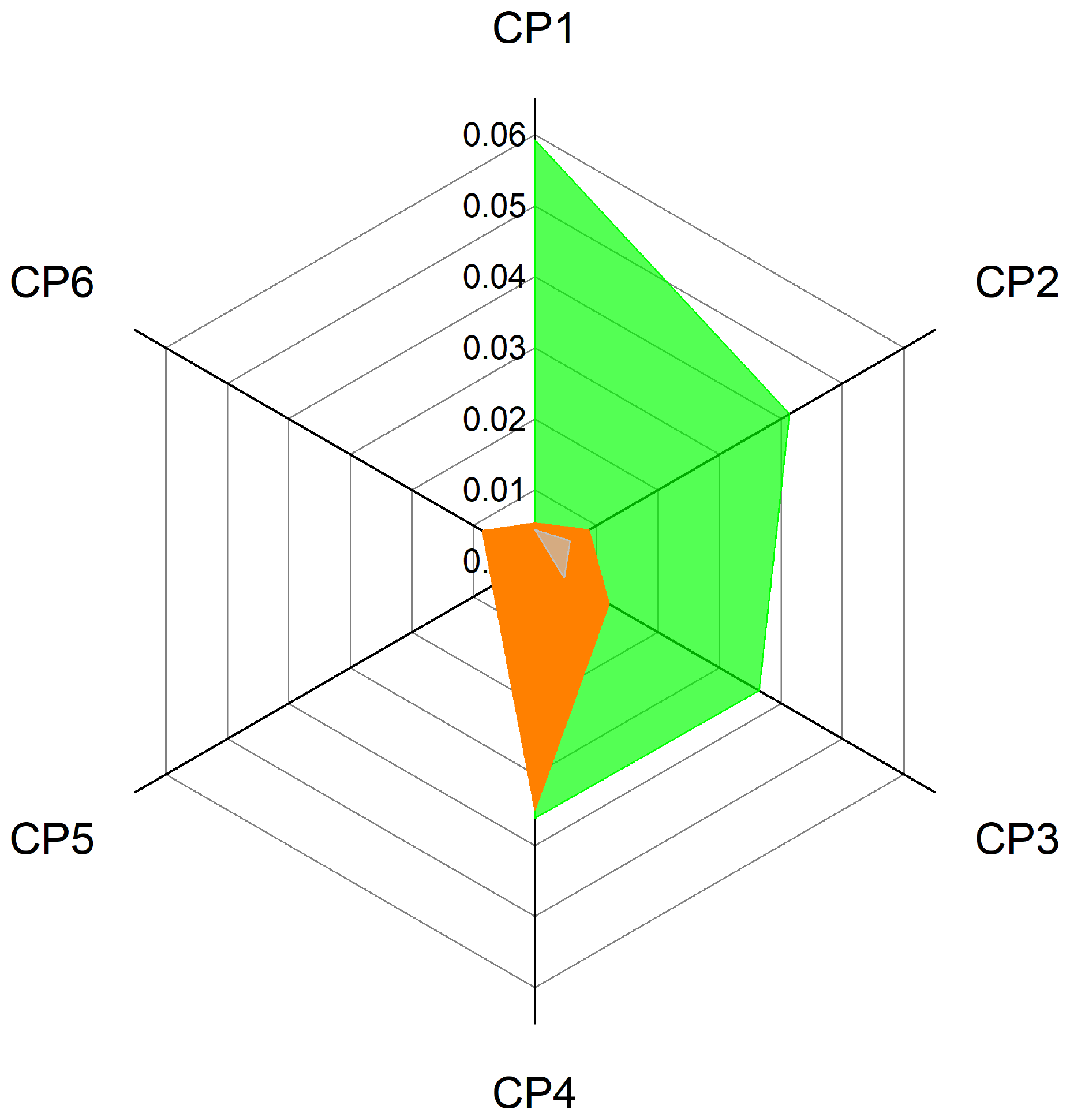}
\label{Fig:Rho_BNNB}}      &
\subfigure[BNNB+M4 ($\nabla
^2\rho$)]{\includegraphics[width=\sizeA]{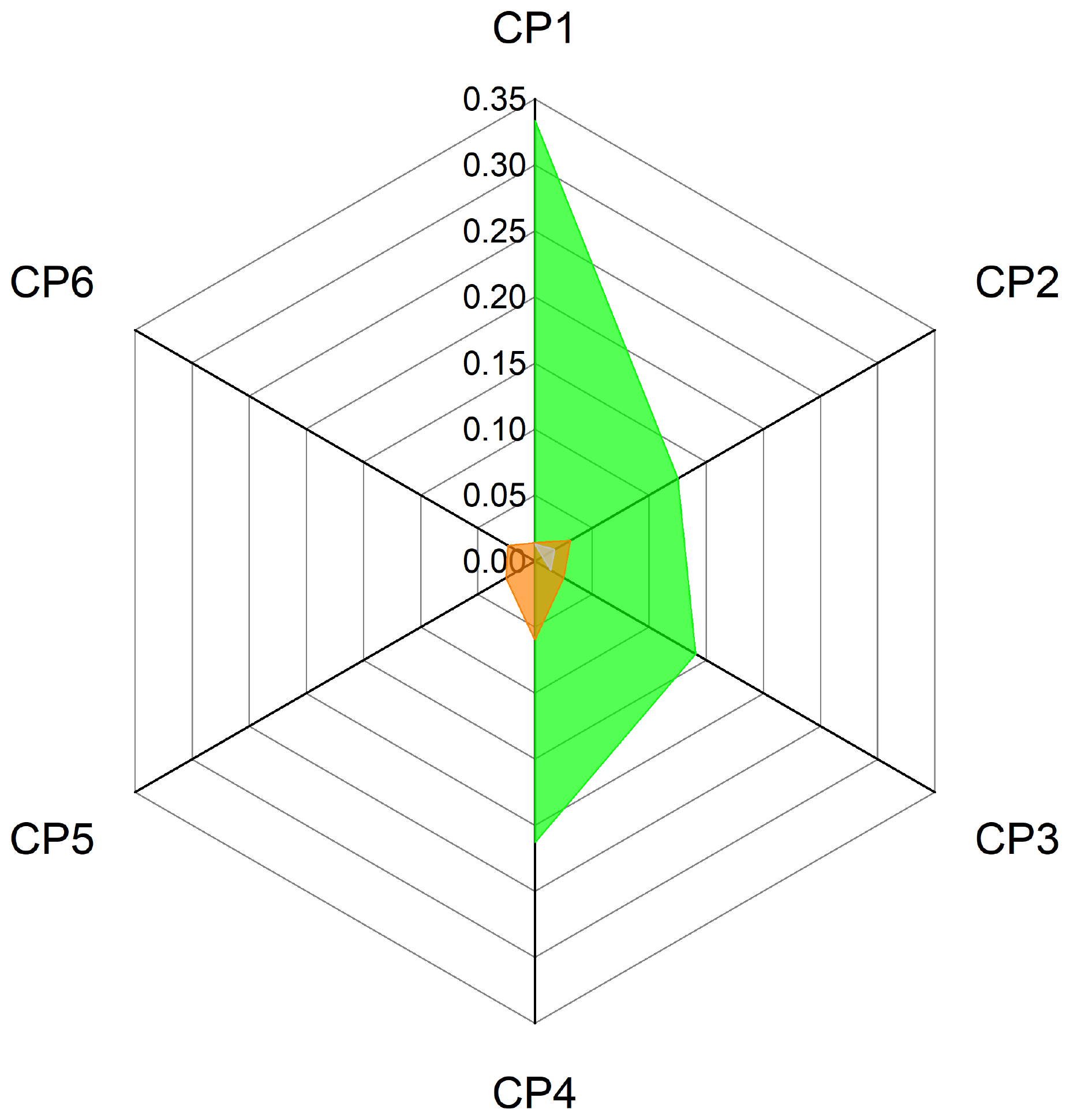}
\label{Fig:Lap_BNNB}}      &
\subfigure[BNNB+M4 (ELF)]{\includegraphics[width=\sizeA]{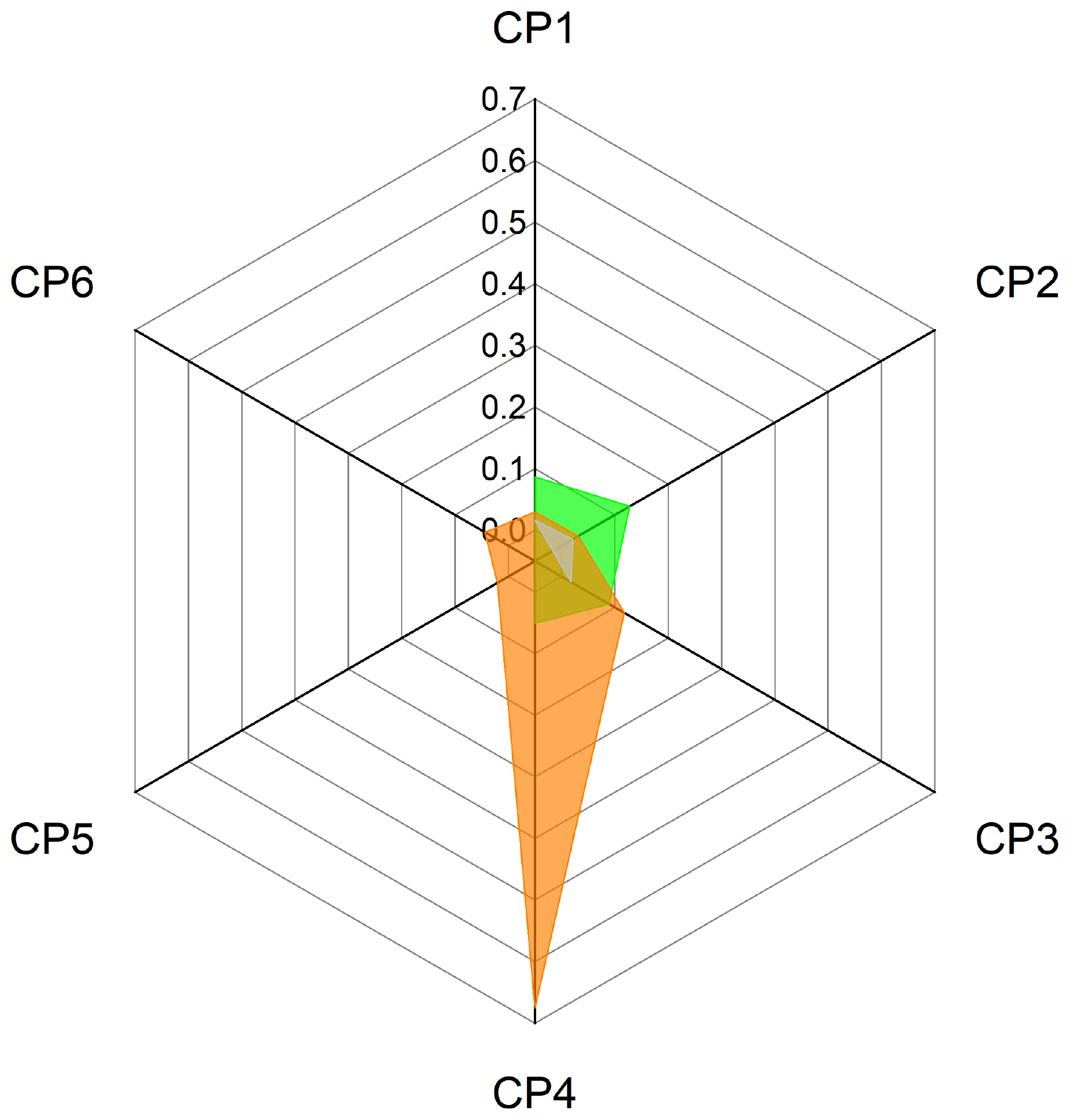}
\label{Fig:ELF_BNNB}}       &
\subfigure[BNNB+M4 (LOL)]{\includegraphics[width=\sizeA]{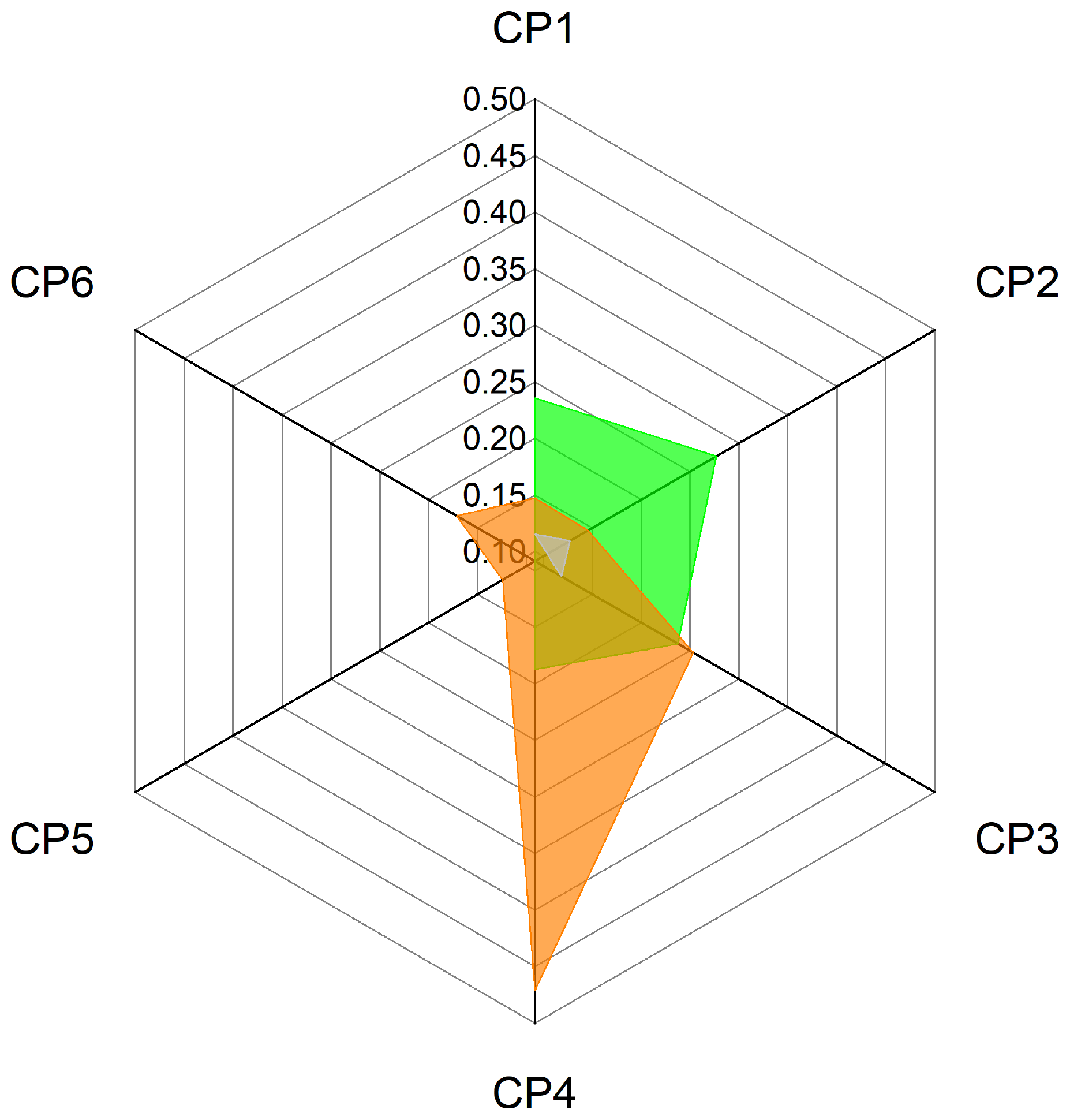}
\label{Fig:LOL_BNNB}}\\
\subfigure[MBNNB+M4 ($\rho$)]{\includegraphics[width=\sizeA]{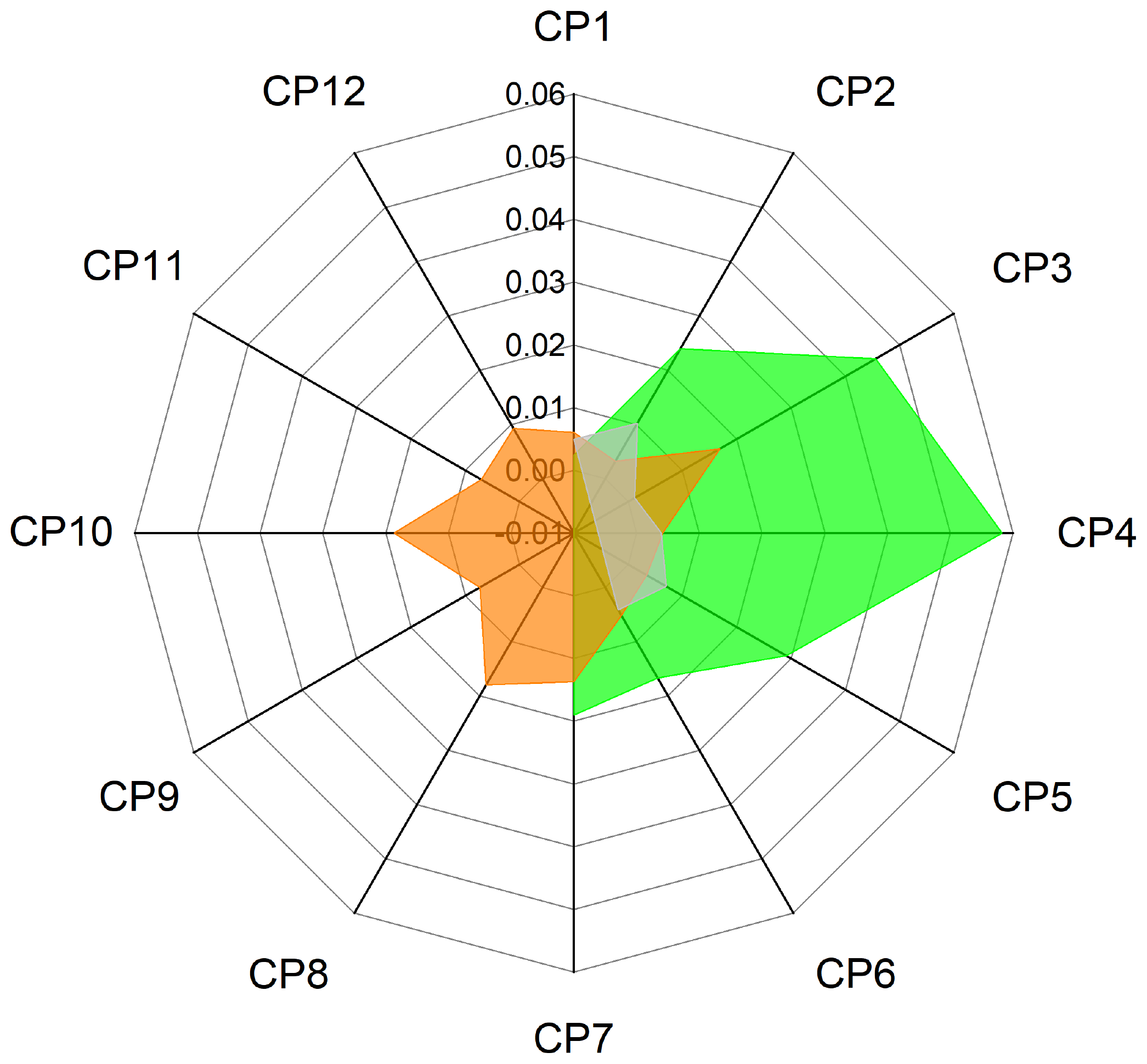}
\label{Fig:Rho_MBNNB}}      &
\subfigure[MBNNB+M4 ($\nabla
^2\rho$)]{\includegraphics[width=\sizeA]{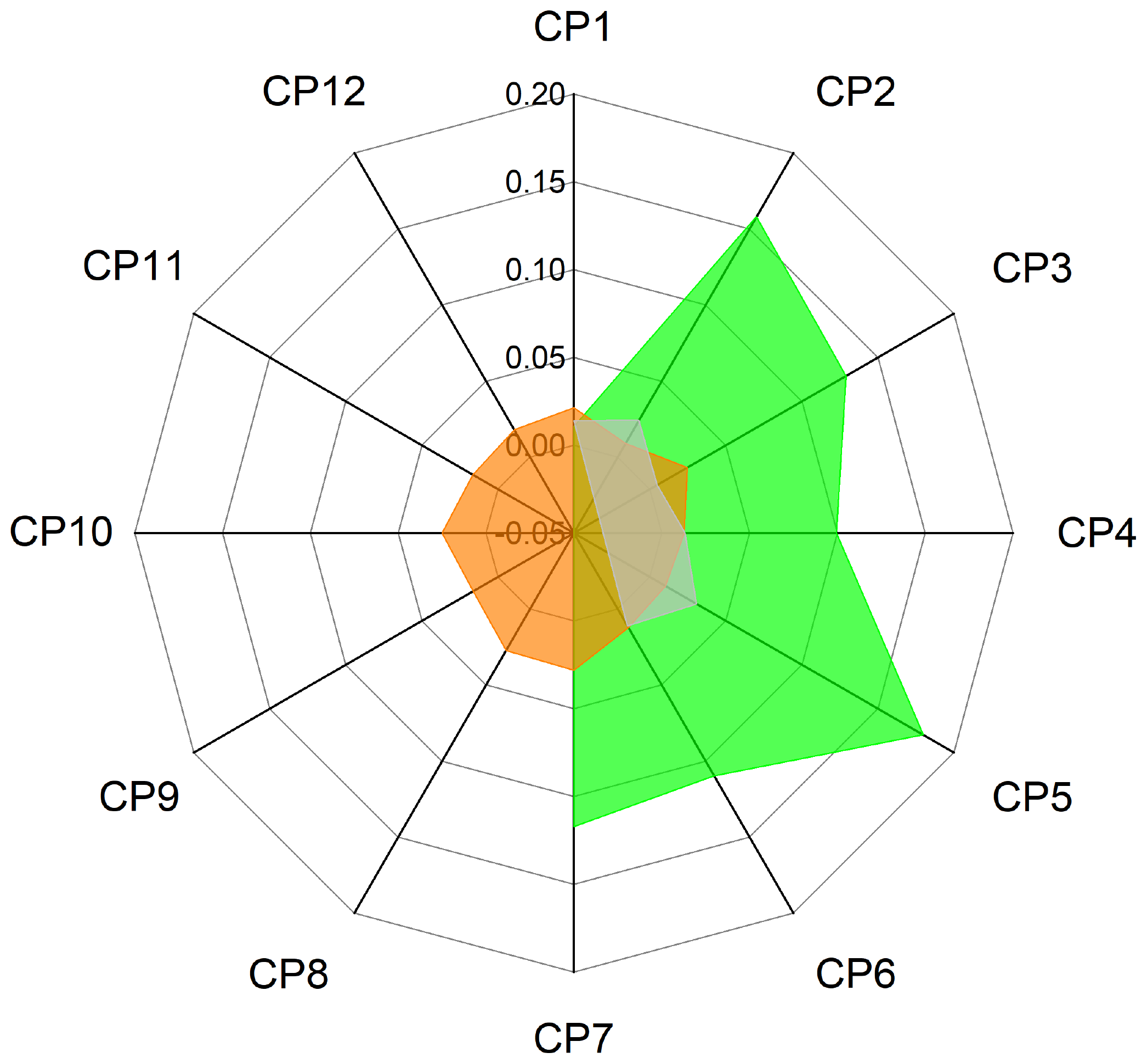}
\label{Fig:Lap_MBNNB}}      &
\subfigure[MBNNB+M4 (ELF)]{\includegraphics[width=\sizeA]{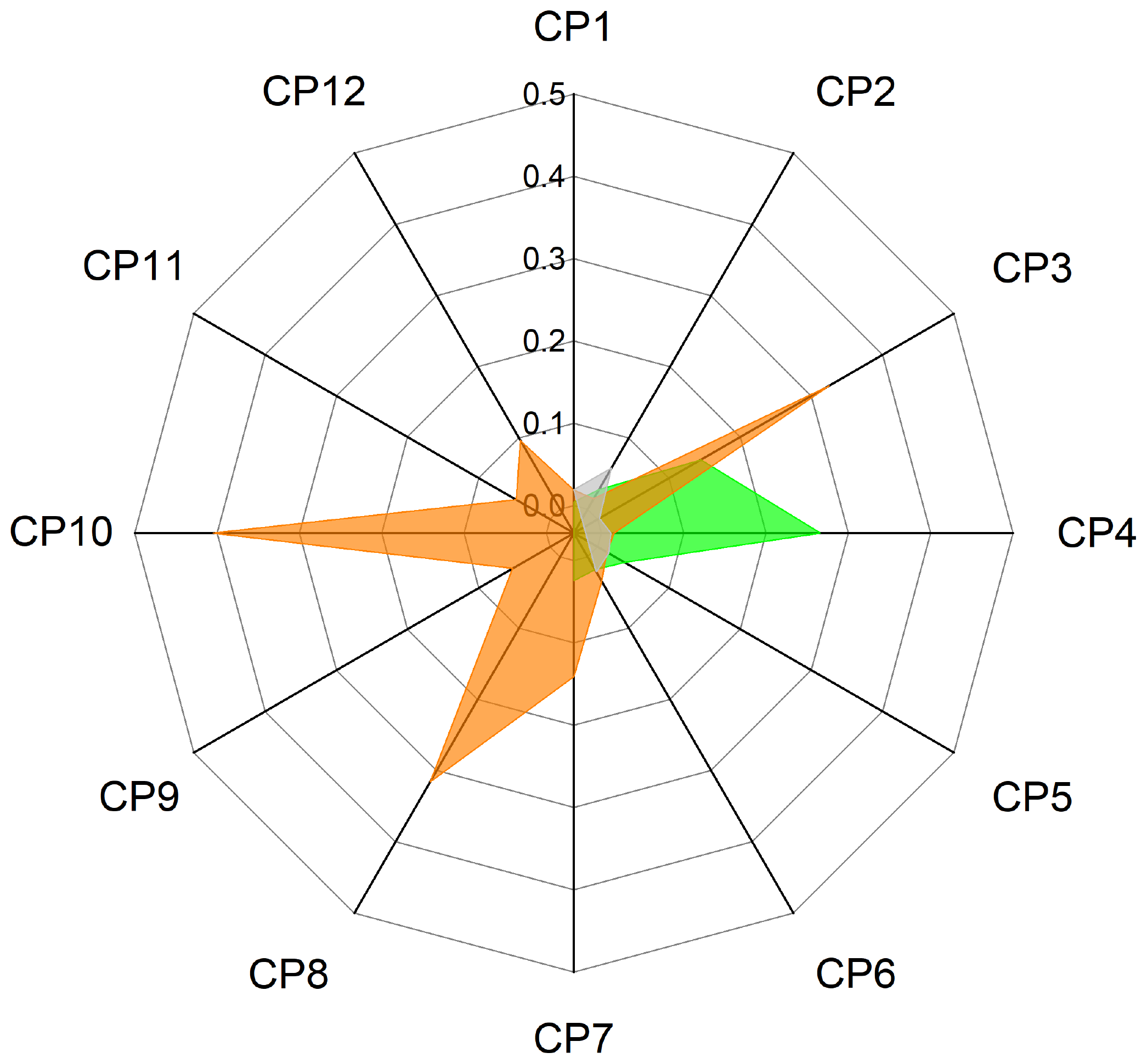}
\label{Fig:ELF_MBNNB}}       &
\subfigure[MBNNB+M4 (LOL)]{\includegraphics[width=\sizeA]{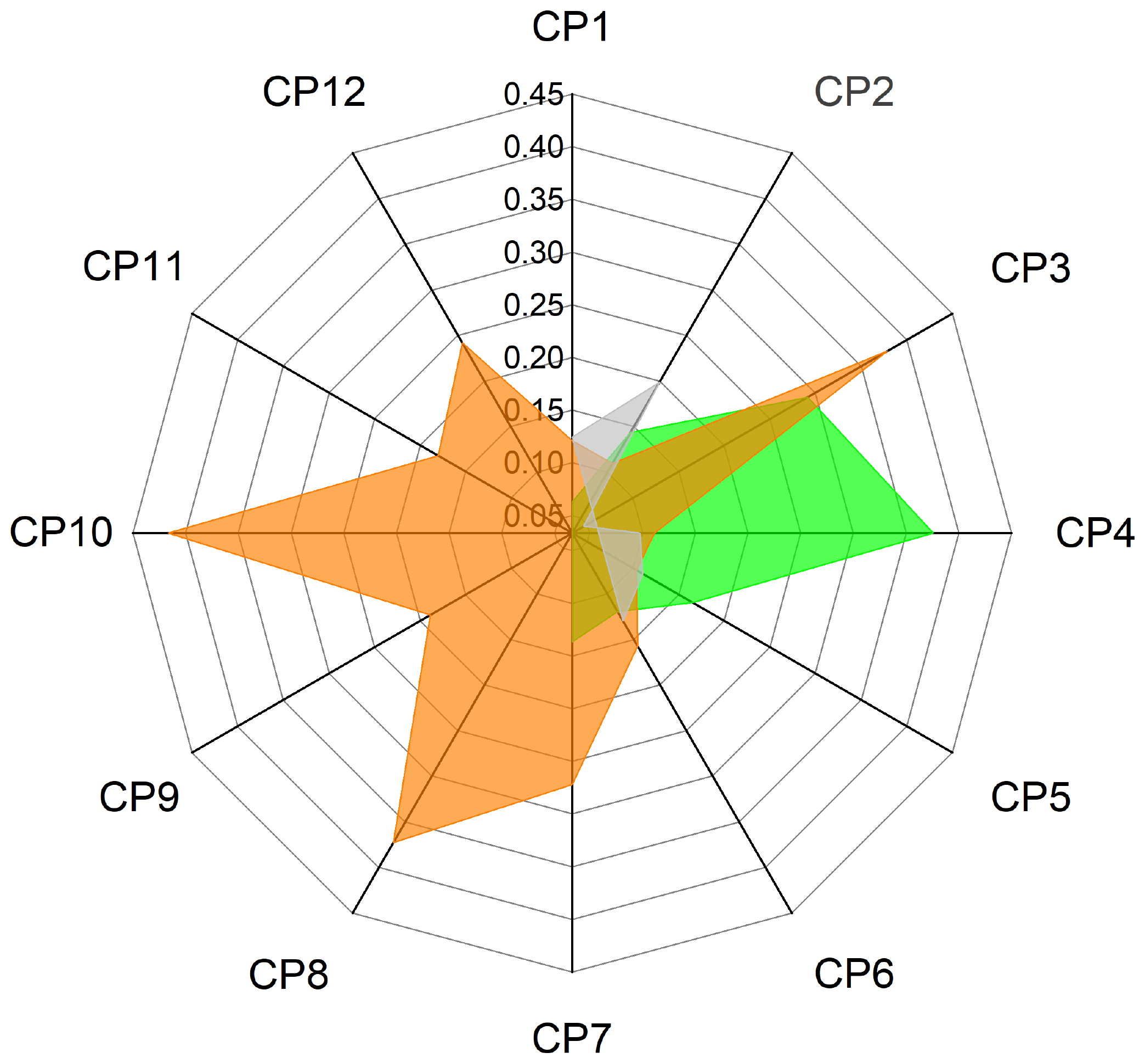}
\label{Fig:LOL_MBNNB}} \\
\end{tabular}
\caption{\label{Fig:Topol} Topology results. The data for Ni clusters is in
green, for Cd clusters is in orange and for Pb clusters is in gray.}
\end{figure}

\section{Conclusions}
\label{Sec:Conclusions}
In this work, the interactions of Ni, Cd, and Pb nanoclusters with
boron-nitride nanobelts with different geometries were studied within the
semiempirical tight binding framework using several methods: best interaction
region, geometry optimization, molecular dynamics, electronic property
calculations, and topology studies.

The more favorable interaction regions were determined using the automated
Interaction Site Screening (aISS) and used as a starting point for the other
simulations. All the structures (individual clusters, nanobelts, and complexes)
were optimized under extreme optimization level to ensure good convergence. The
final energy of the optimized structures was used to calculate the binding
energy for each complex. From the energy analysis, the Ni nanocluster shows the
lower binding energy (most favorable interaction) followed by Cd and Pb
nanoclusters. The final geometry for the MBNNB+Ni2D system shows that the Ni
nanocluster pulled the nanobelt sides in such a way that both sides start
interacting, creating a closed pocket around the metal nanocluster. To study
the complexes' time stability, molecular dynamics simulations were run for a
production time of 100 ps, showing that, in all cases, the heavy metals remain
bound to the nanobelts.

The electronic calculations showed that the topology of the MBNNB changed the
HOMO\-/LUMO distribution when compared to BNNB as there is a break in symmetry
(for BNNB) that induced a redistribution of the orbitals around the twisted
region (for MBNNB). The interaction of heavy metals nanocluster produced a
further modification of the HOMO/LUMO surfaces, being now redistributed around
the region where the metal was located. Due to the metal/nanobelt interaction,
the charges of the nanocluster were also modified to a greater extent for the
Ni nanocluster than for the other metals.

The topological analysis detected the critical points used to better
characterize the type and strength of the interactions. The MBNNB has twice the
number of bond critical points than the BNNB system. This is an indicator that
using MBNNB to adsorb the heavy metals is a better choice. Comparing the values
of the descriptors used (electron density, Laplacian of the electron density,
ELF, and LOL indexes) for all the systems, we can conclude that the Ni
nanocluster is better adsorbed than the Cd and Pb nanoclusters.

Combining the results from the geometry optimization, the binding energy
calculation, and the topological analysis, we can conclude that the Ni
nanoclusters are chemisorbed, whereas Cd and Pb nanoclusters are physisorbed in
both nanobelts, but this adsorption is more favorable for the Möbius
boron-nitride nanobelts.

\section*{CRediT authorship contribution statement}

\textbf{C. Aguiar}: Investigation, Formal analysis, Writing--original draft,
Writing--review \& editing. \textbf{N. Dattani}: Investigation, Resources,
Formal
analysis, Writing--original draft, Writing--review \& editing. \textbf{I.
Camps}:
Conceptualization, Methodology, Software, Formal analysis, Resources,
Writing--review \& editing, Supervision, Project administration.

\section*{Declaration of competing interest}

The authors declare that they have no known competing financial interests or
personal relationships that could have appeared to influence the work reported
in this paper.

\section*{Data availability}
Data will be made available on request.

\section*{Acknowledgements}
We would like to acknowledge financial support from the Brazilian agencies
CNPq, CAPES and FAPEMIG. Part of the results presented here were developed
with the help of a CENAPAD-SP (Centro Nacional de Processamento de Alto
Desempenho em S\~ao Paulo) grant UNICAMP/FINEP--MCT, CENAPAD--UFC (Centro
Nacional de Processamento de Alto Desempenho, at Universidade Federal do
Cear\'a), and Digital Research Alliance of
Canada (via  project bmh-491-09 belonging to Dr. Nike Dattani), for the
computational support.

\newpage

\end{document}